\documentclass[a4paper,11pt]{article}
\pdfoutput=1 

\usepackage{jcappub} 

\usepackage[T1]{fontenc} 

\usepackage{physics}
\usepackage{slashed}
\usepackage{subcaption}
\usepackage{float}
\usepackage[usenames,dvipsnames,svgnames]{xcolor}
\usepackage{multirow}
\usepackage{bm}
\usepackage[shortlabels,inline]{enumitem}
\usepackage[capitalise]{cleveref}
\PassOptionsToPackage{demo}{graphicx}

\def\nl{\nonumber\\}
\usepackage{framed}
\colorlet{shadecolor}{blue!20}
\usepackage[normalem]{ulem}

\makeatletter
\def\@fpheader{\relax}
\makeatother

\title{\boldmath Freeze-in and Freeze-out of Dark Matter with Charged Long-lived Partners}

\author[a]{Sreemanti Chakraborti,}
\author[b]{Victoria Martin}
\author[a]{and Poulose Poulose}
\affiliation[a]{Department of Physics, Indian Institute of Technology Guwahati, Assam 781039, India}
\affiliation[b]{School of Physics and Astronomy, University of Edinburgh, Edinburgh}

\emailAdd{sreemanti@iitg.ac.in}
\emailAdd{victoria.martin@ed.ac.uk}
\emailAdd{poulose@iitg.ac.in}

\abstract{
We present a novel framework capable of addressing the dark matter problem through freeze-in and freeze-out mechanisms, separately or together, depending on the region of the parameter space considered. In the dark matter dynamics, the model features an interplay of thermal production along with sizeable contribution through feeble decay of associated dark fermionic partner, which finally freezes out to the right relic density for a wide range of masses and couplings. Apart from the fermionic dark matter candidate, the model introduces two charged partners, one fermionic and another scalar, which often have delayed decays leading to distinct characteristics of such long-lived particles (LLP) in the colliders like the LHC. Our analysis shows that within the present scenario, LLP of decay length that could be probed at the LHC experiments are compatible with dark matter masses ranging from a few GeV to close to a TeV, as opposed to the requirement of keV-MeV dark matter in simple FIMP scenarios with LLP. In addition, the model presents hitherto unexplored interesting possibilities in the LLP searches, like (i) LLP to LLP to SM cascade decays, which could be searched for within the LHC detectors and (ii) heavy neutral particle decaying within MATHUSLA with two jets and large missing energy. A supplementary aspect of the model is the presence of a heavy neutrino facilitating Type-I seesaw mechanism without disturbing the dark matter side.
}

\begin{document} 
\maketitle
\flushbottom

%
\section{Introduction}
\label{sec:intro}
The presence of dark matter (DM) constituting about 27\%~\cite{Aghanim:2018eyx} of the energy content of the universe is now established beyond reasonable doubt. The observation of cosmic microwave background radiation with tiny anisotropy as measured precisely by the PLANCK experimental collaboration reveals that cold non-baryonic content of the matter density is $\Omega_{cnb}h^2=0.1194\pm 0022$~\cite{Aghanim:2018eyx}. The most studied DM candidates are Weakly Interacting Massive Particles (WIMP), which are thermally produced in the early universe with the relic density surviving after the freeze-out. The attraction of its required annihilation cross section falling in the electroweak ball-park for masses of the order of 100 GeV, therefore providing many natural candidates including the lightest supersymmetric particles~\cite{Jungman:1995df}, are being challenged by the null results of recent direct detection experiments~\cite{Aprile:2017iyp,Akerib:2016vxi,Akerib:2017kat,Cui:2017nnn,Amaudruz:2017ekt,Agnese:2017njq,Agnese:2015nto,Angloher:2015ewa,Arcadi:2017kky,Angloher:2017sxg,Arnaud:2017bjh,Amole:2017dex}. In {simple scenarios of WIMP models}, the nuclear scattering cross section relevant to the direct detection experiments and the annihilation cross section relevant to the relic density are controlled by the same couplings. This include singlet dark matter scenario (please see Ref.~\cite{Athron:2017kgt} for a recent review), the gauged extensions like the Inert Higgs Doublet/Triplet Model~\cite{LopezHonorez:2006gr,Hambye:2009pw,Honorez:2010re,Belyaev:2016lok,Goudelis:2013uca, Poulose:2016lvz, Chao:2012re,Keus:2013hya}, fermionic dark matter models having scalar mediated interaction with the SM sector~\cite{Kim:2006af, Kim:2008pp, Baek:2011aa, Bhattacharya:2015qpa}, and vector dark matter models \cite{Lebedev:2011iq}. A comprehensive summary of these Higgs portal models is presented in a recent review in Ref.~\cite{Arcadi:2019lka}.

This push-and-pull between the two experimental measurements has cornered minimal WIMP scenarios. There are many attempts to alleviate this contest by providing mechanisms for the annihilation process, which are irrelevant to the direct detection process. This includes multipartite dark matter models~\cite{Bian:2014cja,DiFranzo:2016uzc,Bhattacharya:2013nya,Bhattacharya:2016ysw,Ahmed:2017dbb,Bhattacharya:2017fid,Bhattacharya:2013hva,Chakraborti:2018lso,Chakraborti:2018aae,Bhattacharya:2018cgx} with possible conversions from one type to another playing crucial role in achieving the right relic density in specific parameter regions, without affecting the direct detection. The presence of partner particles, having the same symmetry that protects the stability of the dark matter candidate, present in some of these models naturally provides additional annihilation channels which are not relevant to the direct detection. While these attempts are praiseworthy, new directions of thought have emerged providing other equally viable solutions. An attractive possibility is to make the requirement of the annihilation process irrelevant. This is achieved by imagining non-thermal production of the dark matter particles slowly, and gradually building up the required relic density. The mechanism, popularly known as {\emph{freeze-in mechanism}}~\cite{Hall:2009bx} as opposed to freeze-out of the WIMP, needs much smaller couplings, thus the particles are called Feebly Interacting Massive Particles (FIMP)~\cite{Hall:2009bx}. In this case, the direct detection limits are not applicable, as the nuclear scattering cross sections are too small to be probed in the present or near future experiments. However, it may be noticed that the FIMP mechanism works usually with a thermally produced partner particle decaying very slowly to the dark matter particle~\cite{Yaguna:2011qn,Biswas:2016bfo,Biswas:2018aib,Zakeri:2018hhe,DuttaBanik:2016jzv,Tsao:2017vtn}. Such feeble decays make these partner particles live longer than a typical unstable particle. For example, if produced in proton-proton collisions at the LHC, such long-lived particles (LLP) could propagate to perceivable distance before they decay. It provides a possibility for a new way of identifying these scenarios at the collider experiments. Signatures typically include displaced vertex, disappearing charged tracks, and even kinks on the charged tracks. LHC is on the lookout for long-lived particles~\cite{Alimena:2019zri,Sirunyan:2018vlw,Sirunyan:2018njd,Sirunyan:2018pwn,Sirunyan:2018ldc,Sirunyan:2019gut, Khachatryan:2016sfv, Aaboud:2019opc,Aaboud:2019trc,Aaboud:2018hdl,Aaboud:2018kbe,Aaboud:2018aqj,Aaboud:2018arf,Aaboud:2017mpt,Aaij:2017mic,Aaij:2016xmb,Liu:2018wte}. Apart from the dark matter scenarios discussed above, LLP appears in extra-dimensional models~\cite{ArkaniHamed:1998rs,Fairbairn:2006gg}, special scenarios in supersymmetry with very narrow mass splitting between the decaying particle and the final state particles~\cite{Giudice:1998bp,Hewett:2004nw,Banerjee:2016uyt,Nagata:2017gci,Banerjee:2018uut, Harland-Lang:2018hmi}, SM extension with light neutrino mass~\cite{Cottin:2018kmq,Drewes:2019fou} and models explaining baryogenesis~\cite{Cui:2014twa,Choi:2018kto}. The high-luminosity version of the LHC (HL-LHC) has dedicated upgrades keeping in view the possible signatures of the LLP~\cite{CidVidal:2018eel,Drewes:2019fou,Kang:2019ukr,Aboubrahim:2019qpc}. In addition, within the LHC complex a dedicated detector known as (MAssive Timing Hodoscope for Ultra Stable neutraL pArticles) MATHUSLA~\cite{Curtin:2018mvb,Lubatti:2019vkf} is proposed as a large volume surface detector specifically looking for LLP. Detailed study of LLP signatures at the LHC in simplified FIMP models is carried out by~\cite{Belanger:2018sti,Desai:2018pxq,Ghosh2017}, which shows that mass of the dark matter particles should be in the keV - MeV range to be compatible with dark matter relic density, at the same time providing decay lengths of the partner particles that could be probed by LHC.

In this work, we consider a novel scenario where a fermionic dark matter ($\psi$) is produced from the slow decay of a charged fermion ($\chi^+$). 
This necessitates the presence of a charged bosonic degree of freedom, which we consider as a charged scalar $S^+$ for simplicity, allowing $\chi^+$ to decay through $\chi^+ \to \psi~ S^+$.
The Yukawa coupling corresponding to this interaction can be adjusted to obtain the required relic abundance of the dark matter. Considered as an $SU(2)_L$ singlet, the charged scalar requires the presence of a gauge singlet neutral fermion $N$ for it to decay via $S^+ \to N \ell^+$ channel. The Yukawa coupling corresponding to this interaction is independent of the dark sector dynamics. In particular, it is possible that it is small enough to make the charged scalar have sufficiently large decay length to leave distinct LLP signature at the LHC. On the other hand, if the Yukawa coupling in the dark sector is large enough to accommodate thermal production of the dark matter, we have the WIMP scenario with freeze-out relic density. However, one should be careful in dealing with large Yukawa couplings as, in the absence of tree-level diagrams, they can still contribute to the direct detection cross section through photon/$Z$ boson mediated penguin diagrams. 

Typically, one may imagine that when thermal production is enabled, the non-thermal production through the charged fermion decay is not significant. However, an interesting scenario arises when the non-thermal production adds significantly to the number density along with the thermal production. Such a scenario would emerge when the relevant couplings and masses are in favourable ranges. The relic density, in this case, will be dictated by the thermal freeze-out mechanism. Nevertheless, this possibility will lead to more viable regions of parameter space, with consequent effects in other sectors including the collider processes. The result of all these effects provide a feasible mechanism of dark matter of mass in a wide range of a few GeV to close to a TeV, along with LLP that could be probed through the LHC experiments. Presence of multiple LLPs potentially provide another interesting aspect to their search at the LHC through cascade decay of $LLP\to LLP\to SM$, adding a different dimension to the LLP searches. In this article, we shall study the details of the dark matter scenario of the model and indicate the collider possibilities, with the detailed collider phenomenology deferred to a future publication.

We organise this paper with the details of the model given in Section~\ref{Model}, followed by the numerical analysis and discussion of results related to the freeze-in scenario presented in Section~\ref{freezein}, and freeze-out scenario presented in Section~\ref{Freezeout}. In Section~\ref{Collider_discussion} we take a brief look at the collider signatures of the LLP associated with this study, and finally, conclude in Section~\ref{Conclusion}.

%
\section{Model details}
\label{Model}
We consider dark matter to be a gauge singlet fermion denoted by $\psi$, which is made stable by imposing a $Z_2$ symmetry under which it is odd. The non-thermal production of $\psi$ is realised through the slow decay of a partner charged fermion $\chi^+$, which is a vector-like gauge singlet. $\chi^+$ and $\psi$ have the same transformation property under the $Z_2$. This decay necessitates the presence of a charged gauge singlet scalar field, denoted here by $S^+$, which in turn decays to the charged standard lepton and newly introduced gauge singlet vector-like neutral fermion, $N$.  The Yukawa interaction of $N$ with the Standard Model (SM) Higgs field and the left-handed lepton doublet along with its explicit Dirac mass term,  could be made use of to generate the neutrino mass through type-I seesaw mechanism. Keeping this in mind, we introduce three copies of $N$ in our set up. The additional particle spectrum along with their hypercharge and $Z_2$ charge are given in Table~\ref{table:spectrum}. All the SM fields are even under the $Z_2$.

\begin{table}[h]
\centering
{\setlength{\tabcolsep}{1em}
\begin{tabular}{c |c |c }
\hline \hline
{\bf field} & ${\bm Y}$ & ${\bm Z_2}$ \\
\hline\hline
$S^+$ & +2 & + \\
\hline
$N_1, N_2, N_3$ & 0 & + \\
\hline
$\chi^+$ & +2 &$-$ \\
\hline
$\psi$ & 0 & $-$ \\
\hline\hline
\end{tabular}
}
\caption{Additional fields and their charges.}
\label{table:spectrum}
\end{table}

With the above particle content, the Lagrangian of the model is given by
\begin{eqnarray}
\mathcal{L} = &&\mathcal{L} _{SM}+(D_{\mu}S)^{\dagger}(D^{\mu}S)+ \bar{\chi}i\gamma^{\mu}D_{\mu}\chi+\bar{\psi}i\gamma^{\mu}\partial_{\mu}\psi+\sum_i\bar{N_i}i\gamma^{\mu}\partial_{\mu}N_i \nonumber\\
&&- m_{\chi}\ \bar{\chi}\chi - m_{\psi}\ \bar{\psi}\psi -\sum_i m_{N_i}\ \bar{N_i}N_i \nonumber \\
&&- \Big(y_1\ \bar{\chi}S\psi +\sum_i\ y_{2i}\ \bar{N_i}S\ \ell_i  + \sum_i\ y
_{N_i}\ \bar{L_i}\tilde{\Phi}\ N_i+h.c.\Big)\nonumber \\
&&-\big(\mu_S^2\ S^{\dagger}S+\lambda(S^{\dagger}S)^2+\lambda_1\ S^{\dagger}S~\Phi^{\dagger}\Phi\big)
\end{eqnarray}
where $\Phi$ represents the SM Higgs doublet, $L_i$ and $\ell_i$ denote the SM left-handed lepton doublet and right-handed  charged lepton singlet, respectively. The summation index $i$ runs from 1 to 3 indicating the three flavors of leptons. The Yukawa couplings and the mass parameter $m_{N_i}$ are taken to be diagonal, so that the interactions do not lead to any flavor-violating processes. However, the contribution to $g-2$ of the charged leptons arising at the one-loop level are proportional to $y_{2i}^2$, and therefore require to be small~\cite{Jegerlehner:2009ry,Tanabashi:2018oca} for the first two generations, while the $i=3$ it could be more relaxed. Moreover, the presence of neutral fermions $N_i$ can explain neutrino mass generated via Type-I seesaw mechanism. The mass matrix in the interaction basis is
\begin{equation}
\begin{pmatrix}
0&vY_{N}\\
vY^T_N&m_N
\end{pmatrix},
\end{equation}
where $Y_N$ and $m_N$ are $3\times 3$ matrices with $Y_N={\rm diag}\left(y_{N_1}, y_{N_2}, y_{N_3}\right)$ and $m_N={\rm diag}\left(m_{N_1},\right.$ $\left. m_{N_2}, m_{N_3}\right)$. Keeping $m_{N_i}$ at the GeV-TeV scale requires $y_{N_i} \sim 10^{-8}$ to get the right neutrino mass (of $\sim 0.1$ eV). The other Yukawa coupling is limited by perturbativity, and thus we consider $y_1\le \sqrt{4\pi}$. Electroweak symmetry breaking is controlled purely by the Higgs mechanism involving the $\Phi$ field. After the symmetry breaking, the mass of the charged scalar is given by
\begin{equation}
m_{S}^2=\mu_S^2+\frac{\lambda_1v^2 }{2},
\label{mrel}
\end{equation}
whereas $m_\chi$ and $m_\psi$ are free parameters. In addition to the constraints discussed above,  further constraints come from the collider experiments. If kinematically allowed,  $\chi^{+}\chi^{-}$ and $S^{+}S^{-}$ pairs could be produced in $e^+e^-$ collision, leading to dilepton final state accompanied by missing energy. Consequently, to avoid LEP constraints we consider $m_{\chi} \geq$ 45 GeV.  Similarly, the Higgs bosons can decay to $S^+S^-$ pair owing to the quartic coupling, $\lambda_1$, contributing to the Higgs decay. For simplicity, we consider $m_S \geq$ 63.5 GeV to kinematically prohibit this decay. On the other hand, possible contributions to $H\to \gamma\gamma$ and to $H\to \psi\psi$ through the presence of $S^+$  and $\chi^+$ in the loop may need attention.  We discuss the contribution to the di-photon decay channel in Appendix~\ref{sec:hgg}. With $m_S = 150$ GeV, this contribution (Eq.~\ref{eq:hgg}) leads to less than 3\% contribution for $\lambda_1=0.01$. Contribution to the invisible Higgs decay is negligible compared to the standard contributions, as discussed in Appendix~\ref{sec:hpsipsi}. 

Coming to other constraints, the decay pattern of $S\to N\ell$ could look very similar to  supersymmetric slepton decays to neutralinos and leptons, and therefore it is tempting to imagine that the collider constraints from slepton search is applicable to $S^+$. However, the SUSY searches are quite model dependent, and thus not applicable as such in our case. For example, constraints on long-lived stau quoted in Ref.~\cite{CMS:2019twi} is derived within the restricted mass-splitting scenario and within the context of gauge mediated SUSY breaking. Since we have the freedom to choose the couplings and masses within a much larger range, to start with, we do not consider any such collider constraints. However, in the full considerations of collider phenomenology, these constraints should be included.

Turning to the dark matter side, we have different possibilities arising here depending on the kinematics and couplings. First of all, the dominant mechanism for production of the dark matter $\psi$ is the decay of thermally produced charged fermionic partner, $\chi^{+}$. This immediately requires $m_\chi > m_\psi$. The decay proceeds  through their coupling with the singlet charged scalar, $S^+$.  Here, there are two possibilities: (i) when kinematically allowed with $m_\chi\ge m_\psi+m_S$, the decay essentially goes through the on-shell $S^{+}$ ($\chi^{+} \rightarrow \psi S^{+}$); (ii) when $m_\psi+m_N\le m_\chi < m_\psi+m_S$, it results in a three body decay of $\chi^{+} \rightarrow \psi S^* \rightarrow \psi N \ell^{+}$. In the first case, the decay is dictated by a single coupling $y_1$, whereas in the second case it is the product of $y_1y_2$ that matters. As noted above, $y_{2}$ is restricted to be small for the first two generations, while it can be more relaxed for the third generation which couples the $\tau$ lepton with $S^{+}$.
Direct detection experiments can constrain the coupling $y_1$ through processes at one-loop level.   We have discussed the contribution in the Appendix ~\ref{sec:dd}. To have  $\sigma^{\psi\chi}_{\rm SI}\le 10^{-46}$ cm$^2$ as required by the direct detection experiments, for masses of the order of 100 GeV, the coupling  typically comes around $y_1\lesssim 0.05$.

The strength of all these interactions will dictate the relic density of $\psi$, the dark matter candidate. In Table~\ref{table:scenarios} we have listed the distinct possibilities that could arise.

\begin{table}[h]
\begin{center}
\small
\begin{tabular}{l|c|c|l|c}
\multicolumn{3}{l}{Case 1:  $m_\chi > m_S+m_\psi$}\\[2mm]
\hline \hline
Mechanism&$y_1$&$m_\psi$&Possibilities&Viable?\\[1mm] \hline \hline
Thermal&\multirow{2}*{large}&$<m_S$&over-abundance&No\\ \cline{3-5}
production&
&$>m_S$&$\psi\psi\rightarrow S^{+}S^{-}$ enabled freeze-out&Yes \\ \hline
{production through} &small 
&--&freeze-in to saturation&Yes\\[2mm]
{decay of $\chi^+$} & & & \\
\hline \hline
\end{tabular}
\vskip 5mm
\small
\begin{tabular}{l|c|c|c|l|c}
\multicolumn{3}{l}{Case 2:  $m_\psi+m_N < m_\chi < m_S+m_\psi$}\\[2mm]
\hline \hline
Mechanism&$y_1y_2$&$y_1$&$m_\psi$&Possibilities&Viable?\\[1mm] \hline \hline
Thermal&\multirow{2}*{large}&any&$<m_S$&over-abundance&No\\ \cline{3-6}
production&
&restricted
&$>m_S$&$\psi\psi\rightarrow S^{+}S^{-}$ is enabled&Yes \\ \hline
Thermal production&&small&any&freeze-in to saturation&Yes\\ \cline{3-6}
+ production through &small&\multirow{2}*{restricted}&\multirow{2}*{$>m_S$}&$\psi\psi\rightarrow S^{+}S^{-}$ enabled &\multirow{2}*{Yes}\\
\hspace*{4mm}decay of $\chi^+$&
&&&freeze-out& \\[2mm] \cline{4-6}
& & & {$< m_S$} & {over-abundance} & {No} \\
\hline\hline
\end{tabular}
\caption{Distinct scenarios and their viabilities considering relic density of the dark matter.}
\label{table:scenarios}
\end{center}
\end{table}

In the first case with 2-body decay of $\chi^{+}$, the coupling $y_2$ does not affect the scenario. For large values of $y_1$, the dark matter $\psi$ is thermally produced. In addition to the decay of $\chi^+$, the scattering process $S^+S^-\leftrightarrow \chi^+\chi^-$ will also be relevant here. The only way to get the right relic density is by taming the dark matter density through its annihilation. This could be achieved in the present model through the pair annihilation of $\psi$ into $S^{+}S^{-}$ pairs and possible co-annihilations, when kinematically allowed. Note that it is the same coupling which dictates the annihilation process as well. On the other hand, for very small $y_1$, the slow production of $\psi$ gradually builds up the dark matter density.  Apart from $y_1$, the mass splittings would also play crucial role in getting the right relic density.  Moving to the case of 3-body decay of $\chi^+$ as listed in Table~\ref{table:scenarios} Case 2, it is the product of the couplings $y_1y_2$ which is now relevant. Again, for large values of this coupling combination, $\psi$ is thermally produced, and the over-abundance could be tamed with the pair annihilation of $\psi$ for favourable kinematic ($m_\psi > m_S$) and dynamic ($y_1>0.1$, as shown through numerical studies discussed later) conditions. This is quite similar to the large coupling scenario of case 1. On the other hand, for very small values of $y_1y_2$, there are two different situations. One is that $y_1$ is also very small. In that case the only relevant mechanism is the freeze-in mechanism. On the other hand, with large enough $y_1$ thermal production can lead to over abundance when $m_\psi< m_S$, which is tamed through the annihilation process, $ \psi\psi\to SS$ for $m_\psi> m_S$.   In this scenario, when $y_1y_2$ is not too small, {significant contribution} from non-thermal (decay of $\chi$) production of $\psi$ can add to the relic density. However, a proper treatment of this requires solving the coupled Boltzmann equation for both $\psi$ and $S$, as $\psi$ and $S$ may no more be in {chemical} equilibrium. In the present discussion, we shall restrict ourselves to the two limiting cases of either the thermal production or non-thermal production dominate, with the other contribution negligible.

In the following sections we shall discuss the freeze-in and freeze-out scenarios separately, and present a numerical study to understand the specific situations and viable parameter space regions in each of these cases. 

%
\section{Freeze-in solution}
\label{freezein}
The production of DM candidate $\psi$ through the slow decay of gauge produced $\chi^{+}$ gives rise to the freeze-in mechanism by building the DM density to the required value.  The decay of $\chi^{+}$ is controlled by the Yukawa coupling $y_1$ with $S^+$ and $\psi$.   When kinematically possible, it will be a two body decay with $S^+$ produced on-shell. On the other hand, if $m_S+m_\psi\ge m_\chi$, it is a three body decay to final state $\psi N\ell$, mediated by a virtual $S^+$. In the latter case, the decay is dictated by the Yukawa coupling $y_{2}$ of $S^+N\ell^{+}$ along with $y_1$. While in the first case $y_1$ is required to be small enough  for freeze-in to be possible, in the second case the product $y_{2}y_1$ should be of this order.  The relic density depends on the decay width, $\Gamma_{\chi^{+}}$ and the masses of the parent particle and the DM candidate through the relation~\cite{Hall:2009bx}

\begin{equation}
\Omega_{\psi} h^2=\frac{2.19\times 10^{27} g_{\chi}}{g_*^S \sqrt{g_*^{\rho}}}~\frac{m_{\psi}~\Gamma_{\chi^{+}}}{m_{\chi}^2},
\label{eq:relic-freezein}
\end{equation}
where $g_*^{S,\rho}$ are effective degrees of freedom in the bath at freeze-in temperature $T \sim m_{\chi}$. $g_{\chi}$ denotes the number of degrees of freedom corresponding to the decaying particle $\chi^{+}$, which is equal to 2 for a fermion. 

\begin{figure}[h]
\centering
\begin{subfigure}{.3\textwidth}
  \centering
\includegraphics[width=0.7\linewidth]{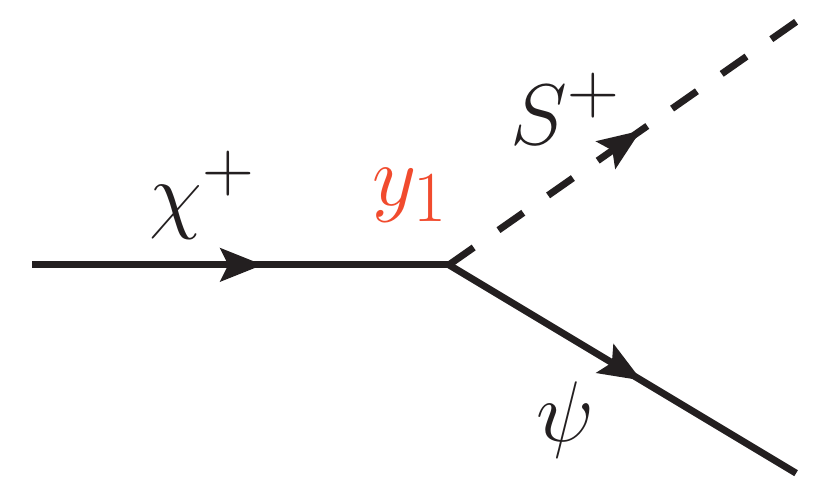}
  \caption{}
  \label{fd2}
\end{subfigure}
\begin{subfigure}{.3\textwidth}
  \centering
  \includegraphics[width=0.7\linewidth]{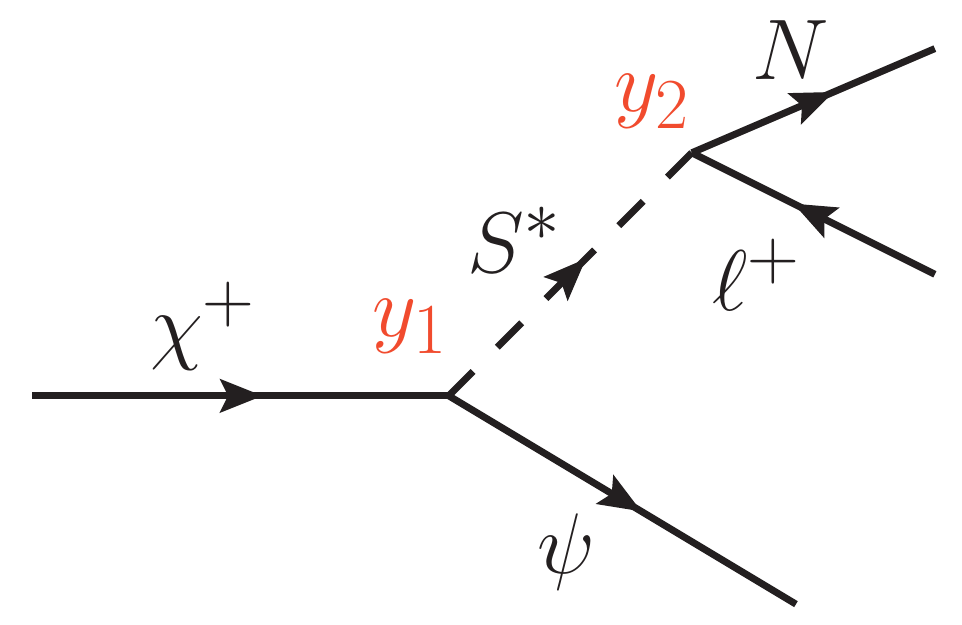}
  \caption{}
  \label{fd3}
\end{subfigure}
\begin{subfigure}{.3\textwidth}
\centering
\includegraphics[width=0.7\linewidth]{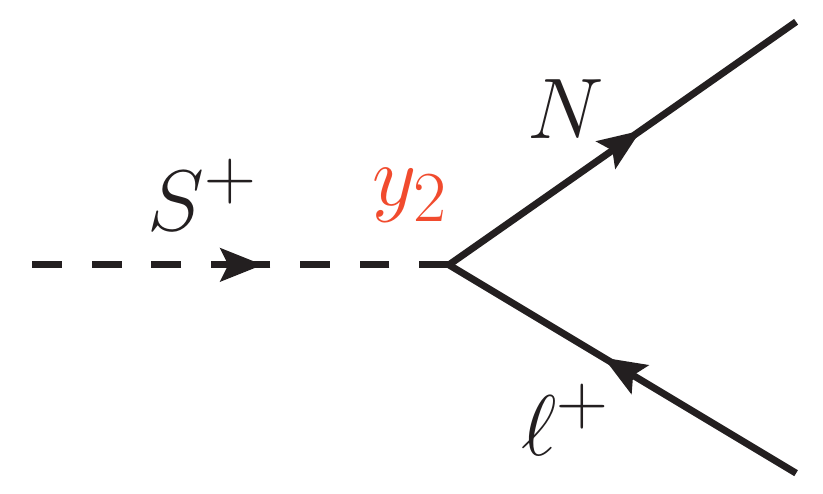}
\caption{ }
\label{fd:Sdecay}
\end{subfigure}
\caption{Feynman diagram showing the two body $(a)$ and three body $(b)$ decays of $\chi^{+}$, and the Feynman diagram showing the decay of $S^{+}$ (c).}
\label{fd:chidecay}
\end{figure}

We shall consider these two distinct cases separately in our numerical analysis below. The charged scalar decays to $N\ell^{+}$, enabled by the Yukawa coupling $y_{2i}$. In the first case of two body decay of $\chi^{+}$, the decay of $S^{+}$ is decoupled from the freeze-in mechanism, and therefore $y_2$ can be independently varied. As mentioned in the introduction, this coupling can potentially induce anomalous $g-2$ for the charged SM leptons, and therefore $y_2$ is required to be small. With small $y_2$, it is possible that $S^+$ has delayed decay, but within the detectors of the LHC experiments. The heavy neutrino $N$ can decay through its coupling with the SM neutrino and the Higgs boson, $y_N$, which is required to be very small to generate the right neutrino mass. Thus, potentially, $N$ could also be long-lived, and mostly decay outside the detector complex giving missing energy signature. In the second case of the three-body decay of $\chi^{+}$, although not decoupled from the relic density formation, $y_2$ can be such that $S^{+}$ has delayed decay to leave distinct signature at the collider detectors. We shall discuss the signatures of this long-lived $S^+$ in the later section. Below we shall focus on the dark matter side of the model and find viable parameter space regions compatible with observations. Feynman diagrams corresponding to the decay of $\chi^{+}$ in the two cases, and that of the $S^{+}$ decay are given in Fig.~\ref{fd2}, \ref{fd3} and Fig.~\ref{fd:Sdecay}, respectively.  We shall discuss these two cases separately below.

\subsection{Two-body decay of  $\chi^{+}$ with $m_{\chi}  >  m_S + m_{\psi}$} \label{freezeincase1}
\begin{figure}[H]
\centering
\includegraphics[scale=0.40]{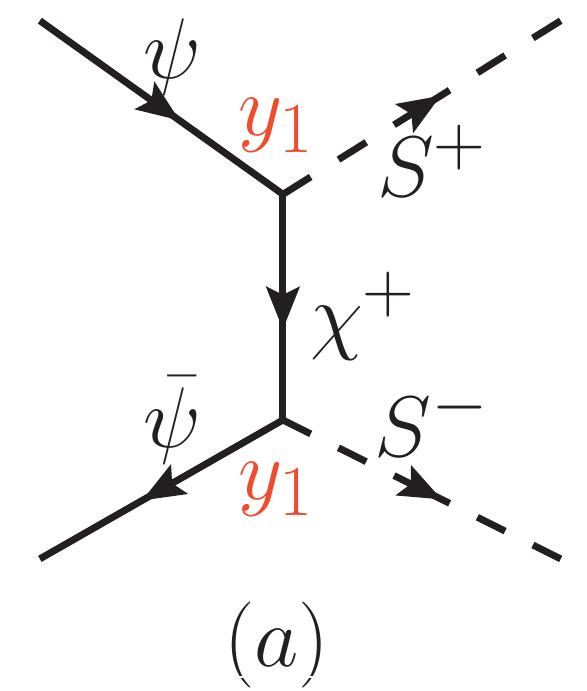}
\includegraphics[scale=0.40]{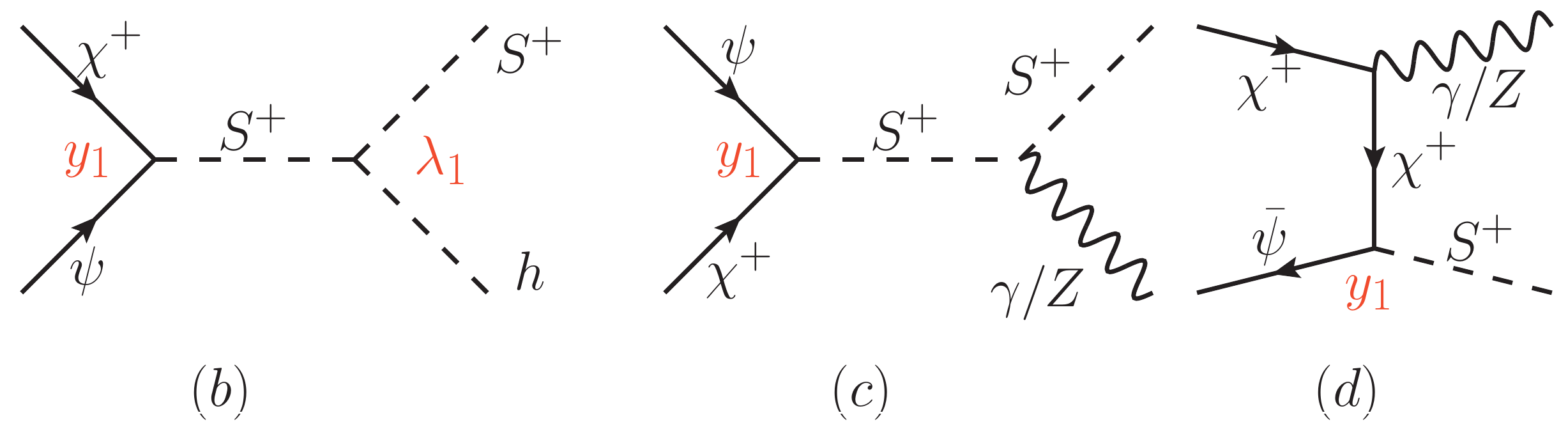}
\caption{Pair annihilation $(a)$ and co-annihilation channels $(b,~c,~d)$ of the dark matter $\psi$.}
\label{fd4}
\end{figure}

In this case, the DM candidate is produced through the two-body decay of the $\chi^{+}$ which is gauge produced abundantly, and in thermal equilibrium.  The decay width of $\chi^{+}$ is given in this case by,
\begin{equation}
\Gamma_{\chi^+ \rightarrow S^+ \psi}=\frac{y_1^2}{16 \pi m_{\chi}^3}\left[\left(m_{\chi}+m_{\psi}\right)^2-m_S^2\right]~
\left[\left(m_\chi^2-m_S^2-m_\psi^2\right)^2-4m_S^2m_\psi^2\right]^\frac{1}{2}
\label{eq_freezein_twobody}
\end{equation}
One may notice that, in principle, a pair of $\psi$ can annihilate into $S^{+} S^{-}$ pairs when kinematically feasible. Such a reaction will be a $t$-channel process mediated by $\chi^{+}$ as shown in the Feynman diagram in Fig.~\ref{fd4}$a$. In addition to the pair annihilation, when the mass splitting is favourable, the co-annihilation processes with Feynman diagrams shown in Fig.~\ref{fd4}$b$,~\ref{fd4}$c$,~\ref{fd4}$d$ could also affect the number density of the dark matter.  However, the cross section of the pair annihilation process is proportional to $y_1^4$, while that of all other  processes are proportional to $y_1^2$, the same coupling that controls the slow production leading to freeze-in mechanism. Thus, the annihilation cross sections are expected to be negligibly small leaving no signature in the relic density.
The relic density for a wide range of parameters is computed using MATHEMATICA with the help of the above relation and compared with the value allowed by PLANCK~\cite{Aghanim:2018eyx} experiment. The relic density in this case depends on the single Yukawa coupling $y_1$, and the masses of the participating particles, $m_\chi$, $m_S$ and the dark matter particle $m_\psi$. We fix the other parameters, which do not affect the relic density as $\lambda=\lambda_1=0.01,~y_N=10^{-8},~y_{2}=10^{-9}$. 

All the three copies of the neutral vector fermion mass are fixed at $m_N=30$ GeV in our scan. While the decay width of $S^{+}$ has a kinematic dependence on $m_N$, the width of $\chi^{+}$ itself is not affected by this parameter. The range of the relevant parameters used in the scan are given in Table~\ref{tb:parameters}. All through we have maintained the mass splitting so that $\chi^{+}$ can undergo a two body decay. Scanning over 50,00 random points, we selected those satisfying the right relic density, and analysed those for possible correlations and constraints.
\begin{table}[h]
\begin{center}
\begin{tabular}{c |c }
\hline\hline
  1 GeV $ \le m_{\psi} \le $ 1000 GeV  &  65 GeV $\le m_S \le $ 1000 GeV  \\ \hline
  $m_S+m_\psi \le m_{\chi} \le $ 1 TeV &   $10^{-14} \le y_1 \le 10^{-8}$  \\ \hline\hline
\end{tabular}
\caption{Range of relevant parameters considered for the numerical scan in this study.}
\label{tb:parameters}
\end{center}
\end{table}

\begin{figure}
\centering
\includegraphics[width=0.48\textwidth]{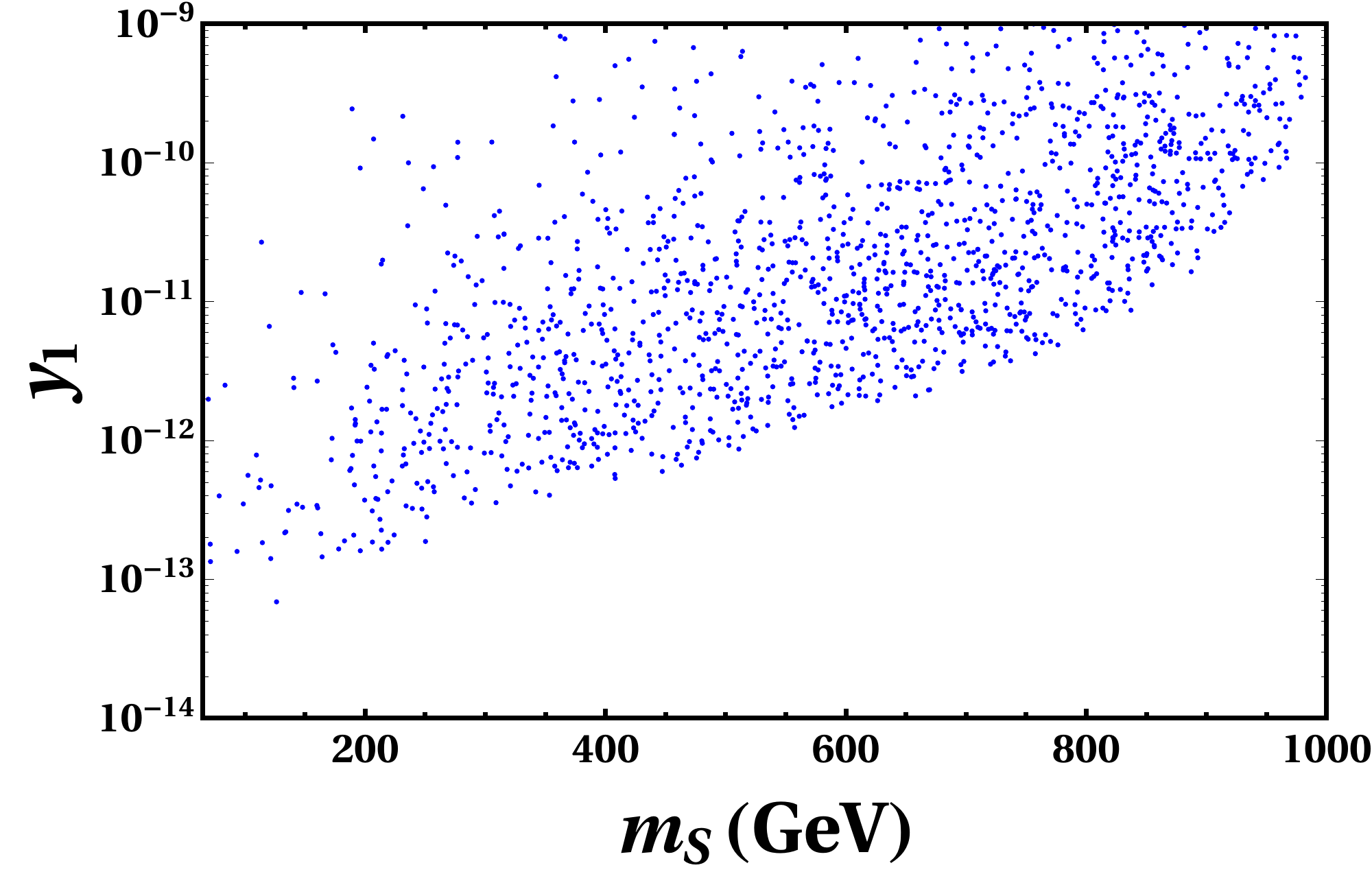}
\includegraphics[width=0.48\textwidth]{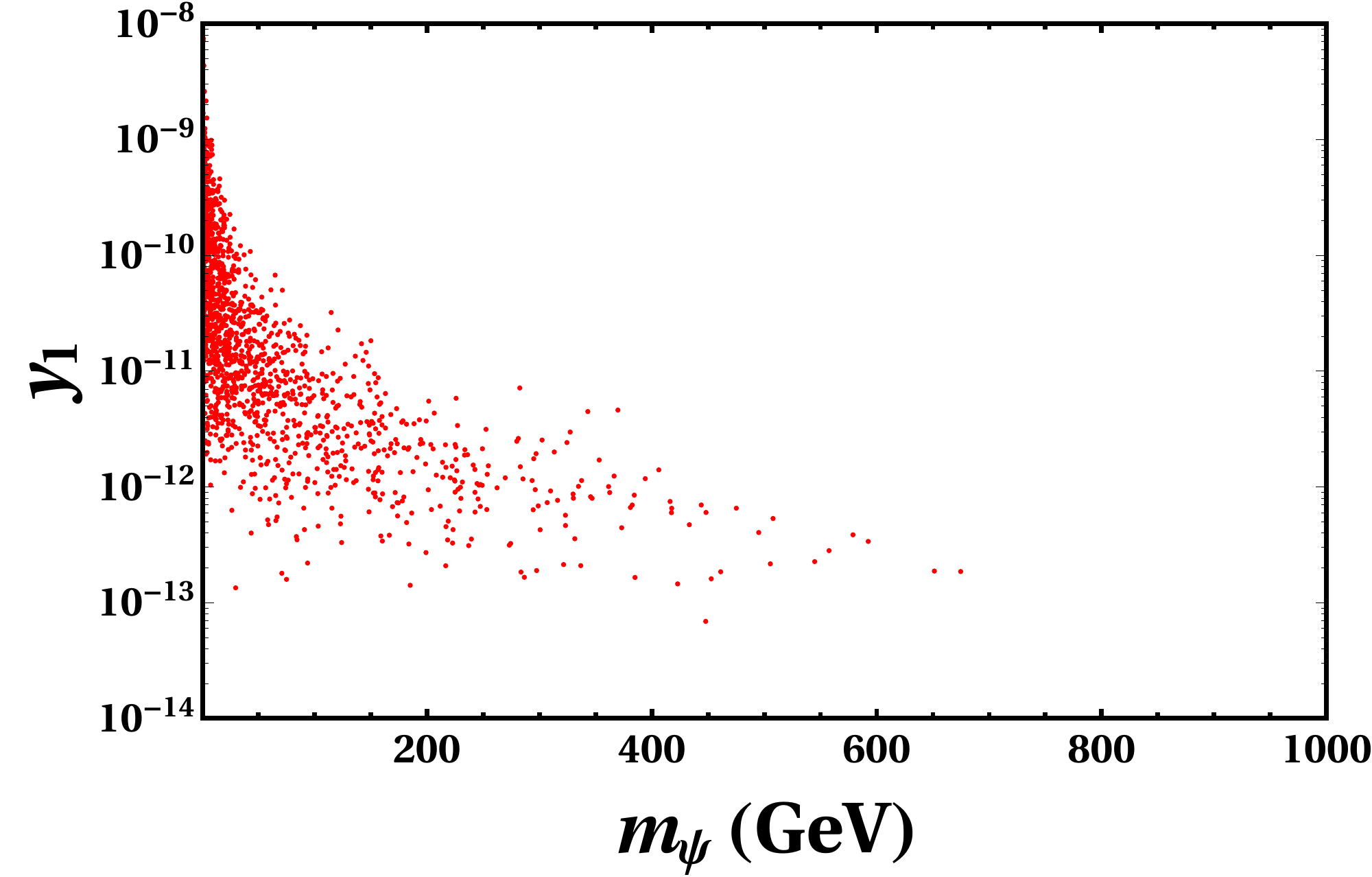}\\
\includegraphics[width=0.48\textwidth]{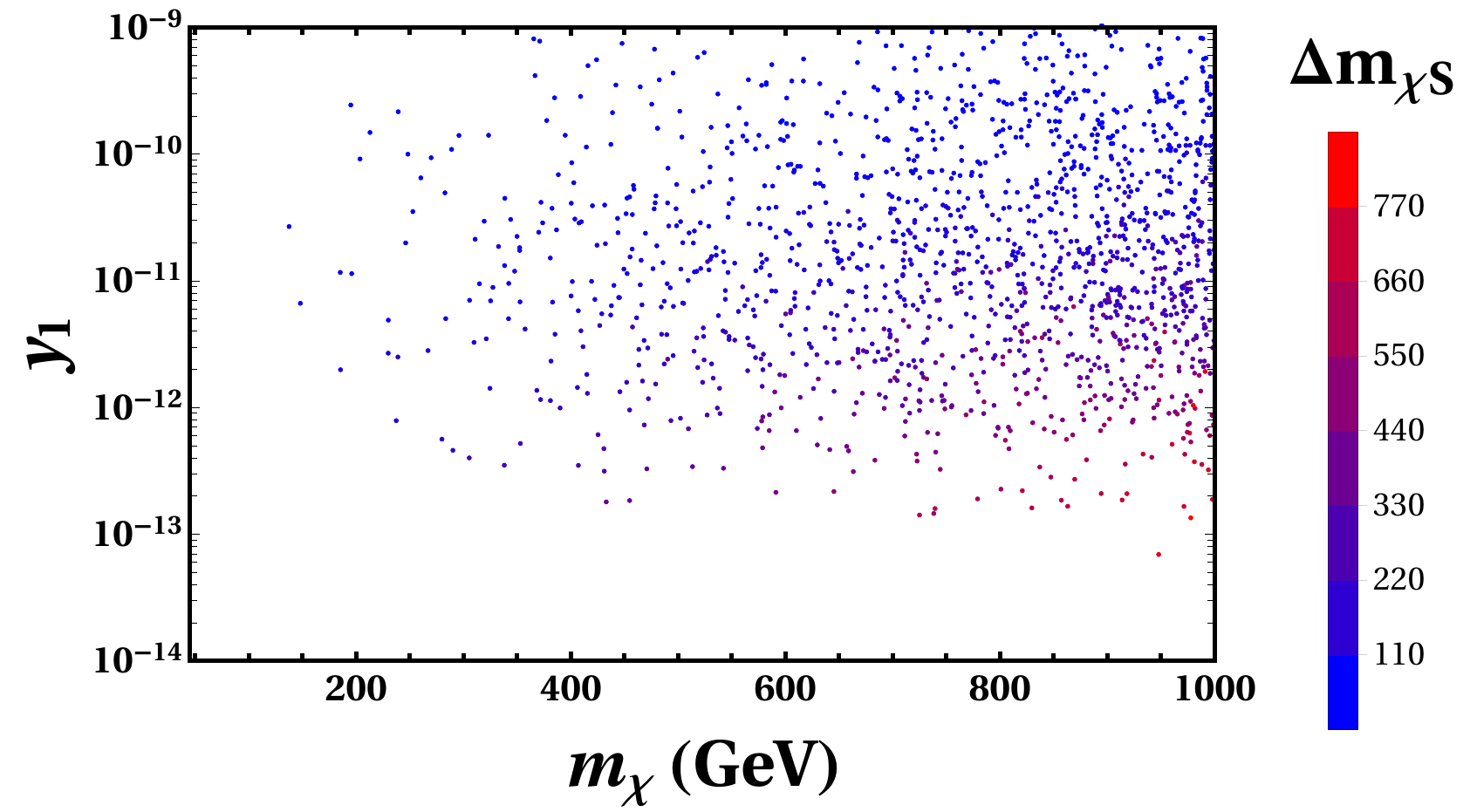}
\includegraphics[width=0.48\textwidth]{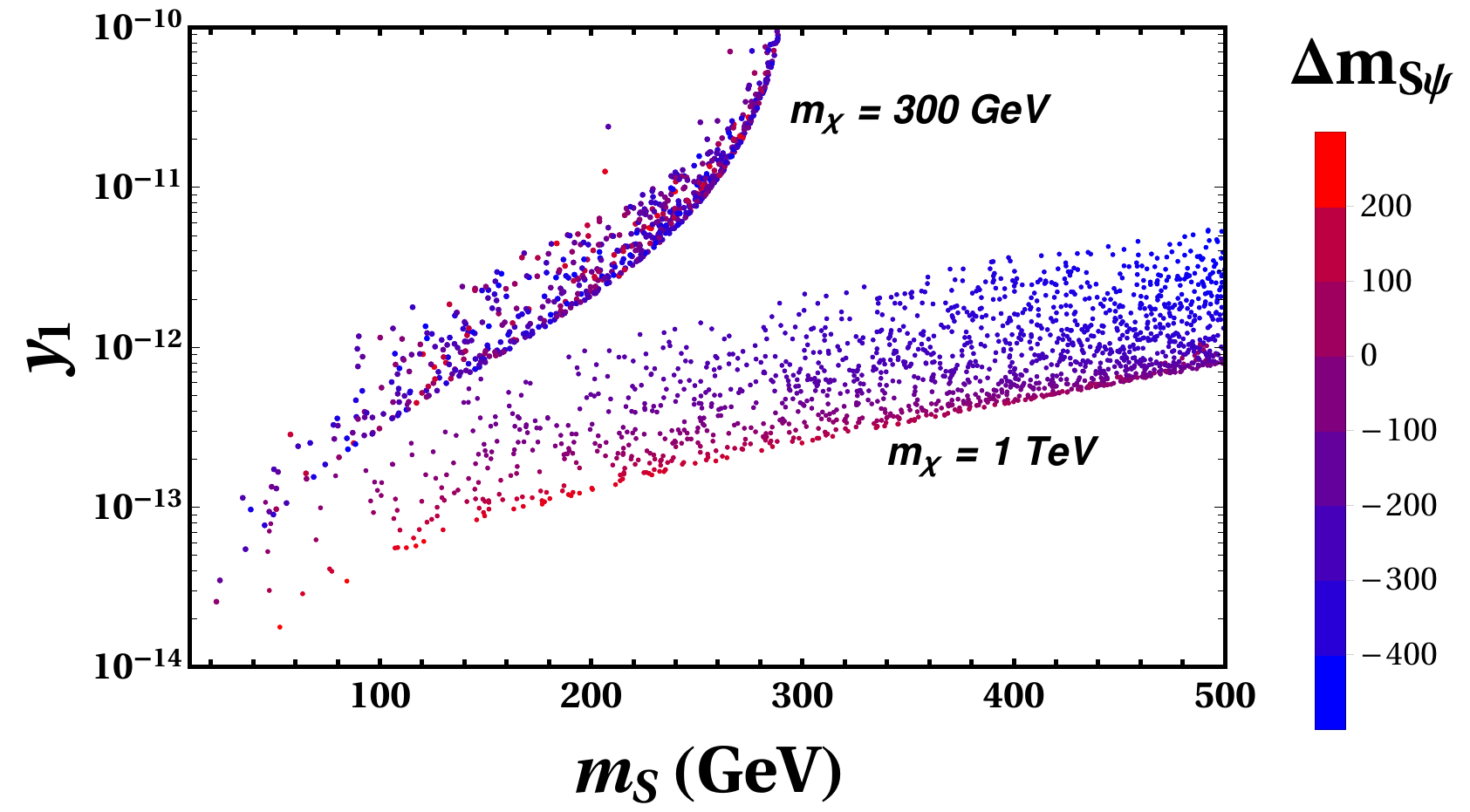}
\caption{Yukawa coupling $y_1$ vs.~mass of the non-SM particles that satisfy the relic density bound. 
In plots in the top row and bottom-left $m_S,~m_{\psi}$ and $m_{\chi}$ are varied from a few GeV to 1 TeV. Other parameters are as per Table~\ref{tb:parameters}. In the bottom-right plot, $m_\chi=300$ GeV and 1 TeV are considered, while $m_\psi$ is varied in the full range. }
\label{case1_freezein}
\end{figure}

In Fig.~\ref{case1_freezein} the correlation between $y_1$ and relevant masses are given.  As seen from $y_1$ vs.~$m_S$  larger masses require larger couplings to compensate for the effect of the phase pace. On the other hand, in the case of $y_1$ vs.~$m_\psi$, the phase space suppression for larger $m_\psi$ is smaller compared to the enhancement in the invariant amplitude, as is clear from the decay width expression in Eq. \ref{eq_freezein_twobody}. Further, there is an additional factor of $m_\psi$ in the expression for $\Omega_{\psi}\rm h^2$ in Eq.~\ref{eq:relic-freezein}, pulling down the decay width to maintain the same relic density. The correlation between the parent particle mass ($m_\chi$) and $y_1$  clearly indicates that $y_1$ is preferred to be in the range of $10^{-13} - 10^{-9}$ to satisfy the required relic density.  As expected, smaller mass splitting $\Delta m_{\chi S}=m_\chi-m_S$ requires larger $y_1$  to compensate for the narrow phase space available. The right-bottom plot in Fig.~\ref{case1_freezein} shows $y_1$ against $m_S$ for two different $m_\chi$ values of 300 GeV and 1 TeV. The mass splitting $\Delta m_{S\psi}=m_\psi-m_S$ is indicated in different colours. While in the case of lighter $m_\chi$ the mass splitting is not very relevant, in the case of heavier $\chi^{+}$ smaller $y_1$ is preferred for larger $m_\psi$ values, in agreement with top-right plot and  the discussion above.

\subsection{Three body decay of $\chi^{+}$ with  $m_\psi+m_N+m_\ell < m_{\chi}  <  m_S + m_{\psi}$}
\label{freezeincase2}
In this case, the decay of $\chi^{+}$ to on-shell $S^{+}$ is not kinematically possible. However, three body decay enabled through a virtual $S^{+}$ decaying into $N\ell$ pair, as illustrated in the Feynman diagram in Fig.~\ref{fd:chidecay} (b), produces the dark matter. The decay width $\Gamma_{\chi^+ \rightarrow \psi N l^+}$ is proportional to $(y_1 y_2)^2$, as given by
\begin{eqnarray}
\Gamma_{\chi^+ \rightarrow \psi N \ell^+}&= \displaystyle\frac{(y_1y_2)^2}{256 \pi^3\, m_{\chi}^3}\displaystyle\int_{X_{\mathrm{min}}}^{X_{\mathrm {max}}}&\mkern-18mu dX
\left[(m_{\chi}^2-X-m_{\psi}^2)^2-4Xm_{\psi}^2\right]^\frac{1}{2}\left[(X-m_N^2+m_\ell^2)^2-4Xm_\ell^2\right]^\frac{1}{2} \nonumber \\
&&\times ~~\frac{\left[(m_{\chi}+m_{\psi})^2-X\right]\left[X-m_N^2-m_\ell^2\right]}{X~\left[(X-m_S^2)^2+\Gamma_S^2m_S^2\right]},
\end{eqnarray}
where \(X_{\mathrm{max}}= (m_{\chi}-m_{\psi})^2 \) and \(X_{\mathrm{min}}= (m_N+m_\ell)^2.\) Having the product of the couplings controlling the production, unlike in the previous case, the following distinct possibilities arise in this case, depending on the mass splitting $\Delta m_{\psi S} = m_\psi - m_S$.
\begin{enumerate}
\item $y_1y_2\ll 1$, $y_1\ll 1$ for any $\Delta m_{\psi S}$:
In this case, the coupling $y_1$ being small { $(\le 10^{-8})$}, the annihilation channel $\psi \psi \rightarrow S^{+}S^{-}$ is not effective leading to a situation similar to Case~1 described in Section~\ref{freezeincase1}. 
$\psi$ produced via 3-body decay of $\chi^{+} \rightarrow \psi N \ell^{+}$ slowly saturates to leave the required relic density. We shall show that, for a large range of mass of $S^{+}$, it is possible to find couplings compatible with the freeze-in scenario in the range of $10^{-11} \le (y_1y_2) \le 10^{-5}$. 

\item $y_1y_2\ll 1$,  $y_1\sim 1$, with $\Delta m_{\psi S}< 0$:  In this case the production is mostly thermal through $SS\to \psi\psi$ with possible addition from the decay of $\chi^+$, the contribution being dependent of the value of $y_1y_2$.  Since the annihilation $\psi \psi \rightarrow S^{+}S^{-}$ is kinematically disfavoured, it results in over-abundance. The co-annihilation channels could, however, be relevant, as the kinematics may not restrict those.

\item $y_1y_2\ll 1$,  $10^{-8}<y_1\sim 1$, with $\Delta m_{\psi S}>0$: In this case again, the thermal production through $SS\to \psi\psi$ and from the decay of $\chi^{+}$ builds up the number density of the dark matter particle. However, owing to sizeable $y_1$ the now kinematically allowed annihilation cross section is significant, leading to favourable relic density through freeze-out. The resultant relic density at the saturation depends on various factors including the rate of decay of $\chi^{+}$ and the annihilation cross section of $\psi \psi \rightarrow S^{+}S^{-}$ channel. 
\end{enumerate}
We shall not discuss case 2 above any further, as this does not lead to the right relic density. The other two cases, case 1 where the thermal production and annihilation are not relevant, owing to small value of $y_1$, and case 3 with large $y_1$ are discussed with numerical results below.

\noindent\emph{\textbf{In the absence of annihilation:}} We shall first consider the situations 1 listed above, where the annihilation does not play a role. Scanning the parameters within the range specified by Table~\ref{tb:parameters2}, we explore the parameter space providing the right relic density.
\begin{table}[h]
\begin{center}
\begin{tabular}{c |c }
\hline\hline
  1 GeV $ \le m_{\psi} \le $ 1000 GeV  &  65 GeV $\le m_S \le $ 1000 GeV  \\ \hline
 $m_N+m_\psi \le m_{\chi} \le $ 1 TeV &   $10^{-12} \le y_1y_2 \le 10^{-5}$  \\ \hline\hline
\end{tabular}
\caption{Range of relevant parameters considered for scan. The condition, $m_\chi < m_\psi+m_S$ to kinematically disable $\chi^{+} \rightarrow S^{+}\psi$ with $S^{+}$ on-shell is further imposed. A few selected $m_N$ values are considered for the analysis.}
\label{tb:parameters2}
\end{center}
\end{table}

\begin{figure}
\centering
\includegraphics[width=0.48\textwidth]{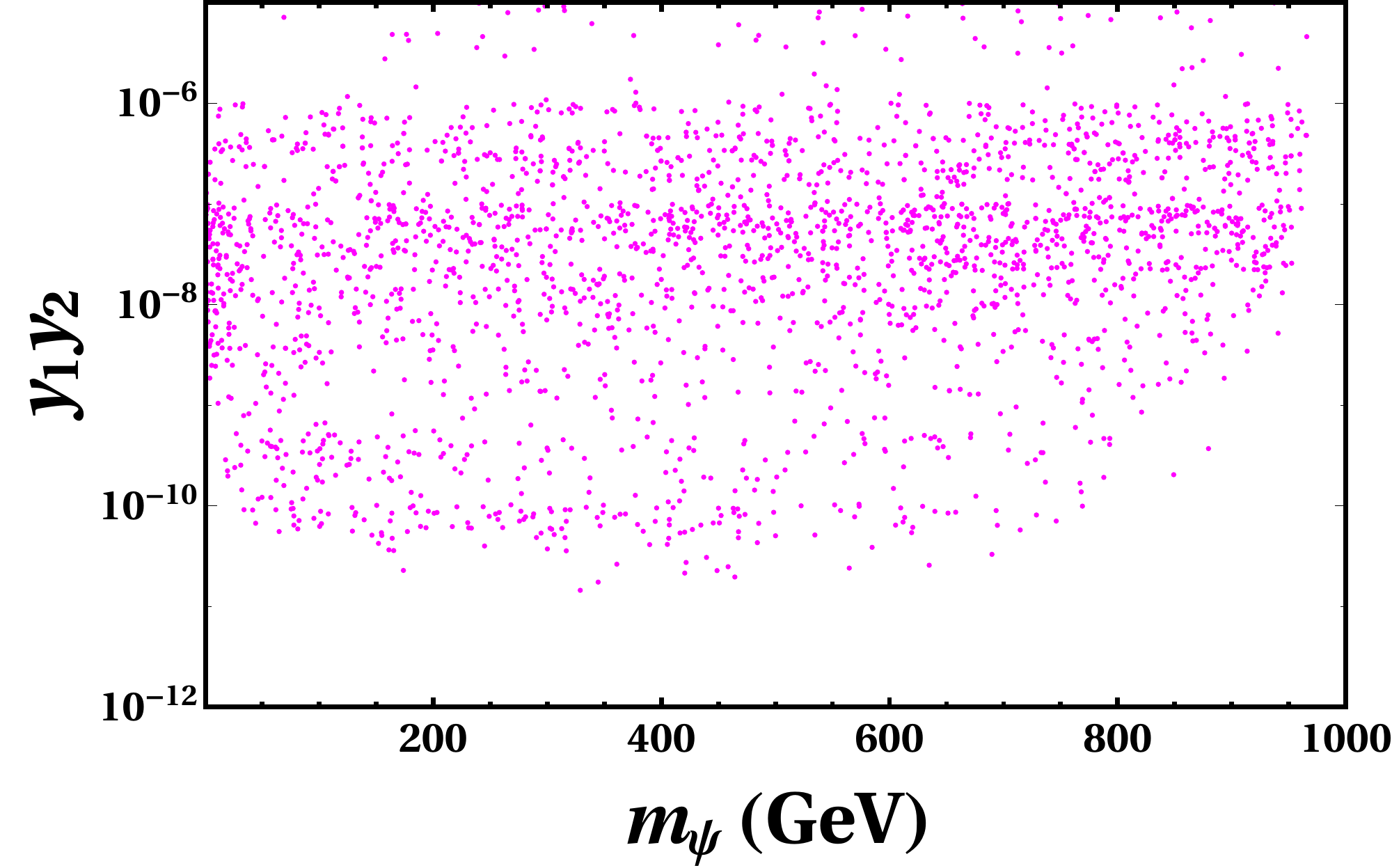}
\includegraphics[width=0.48\textwidth]{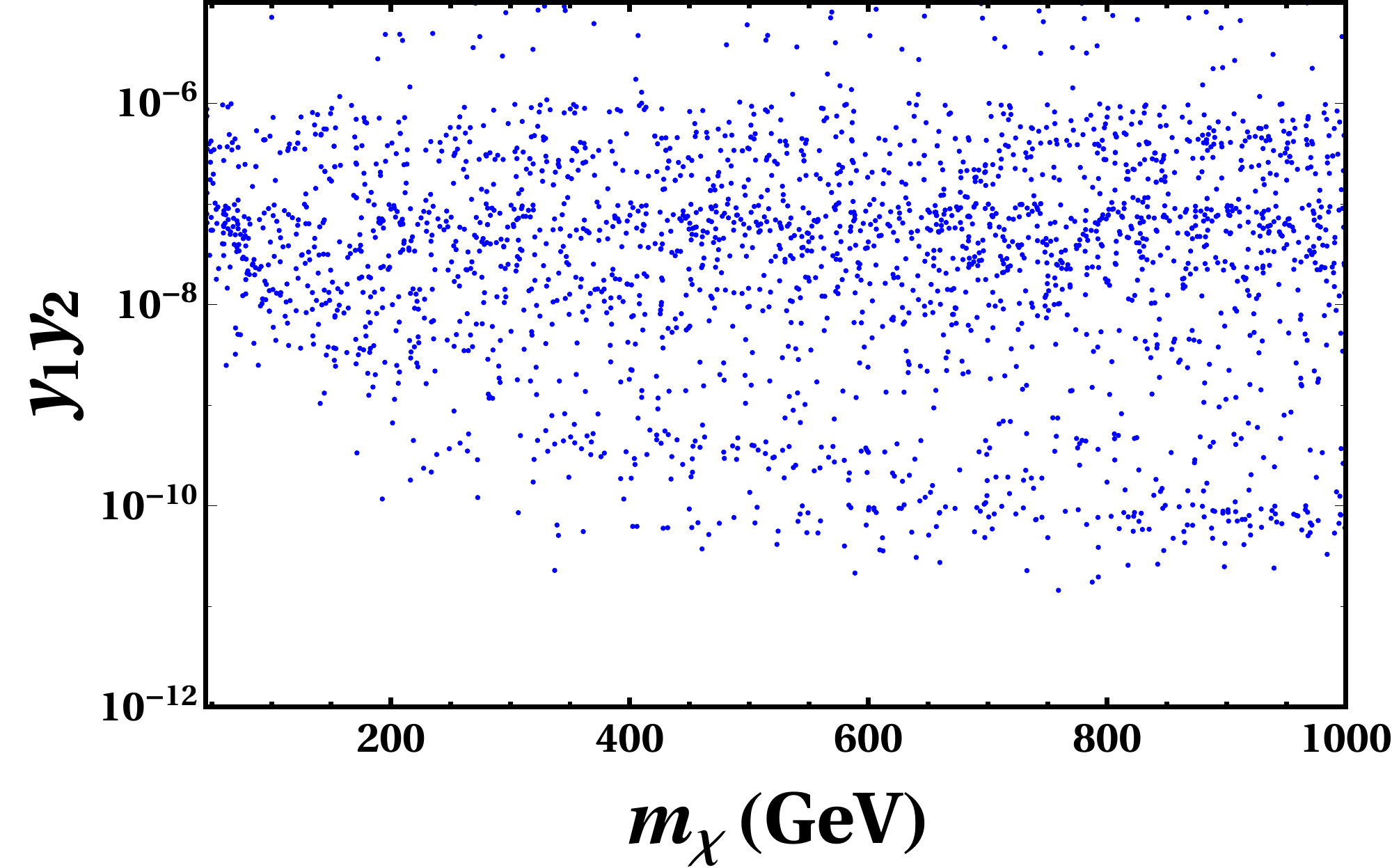}\\[4mm]
\includegraphics[width=0.48\textwidth]{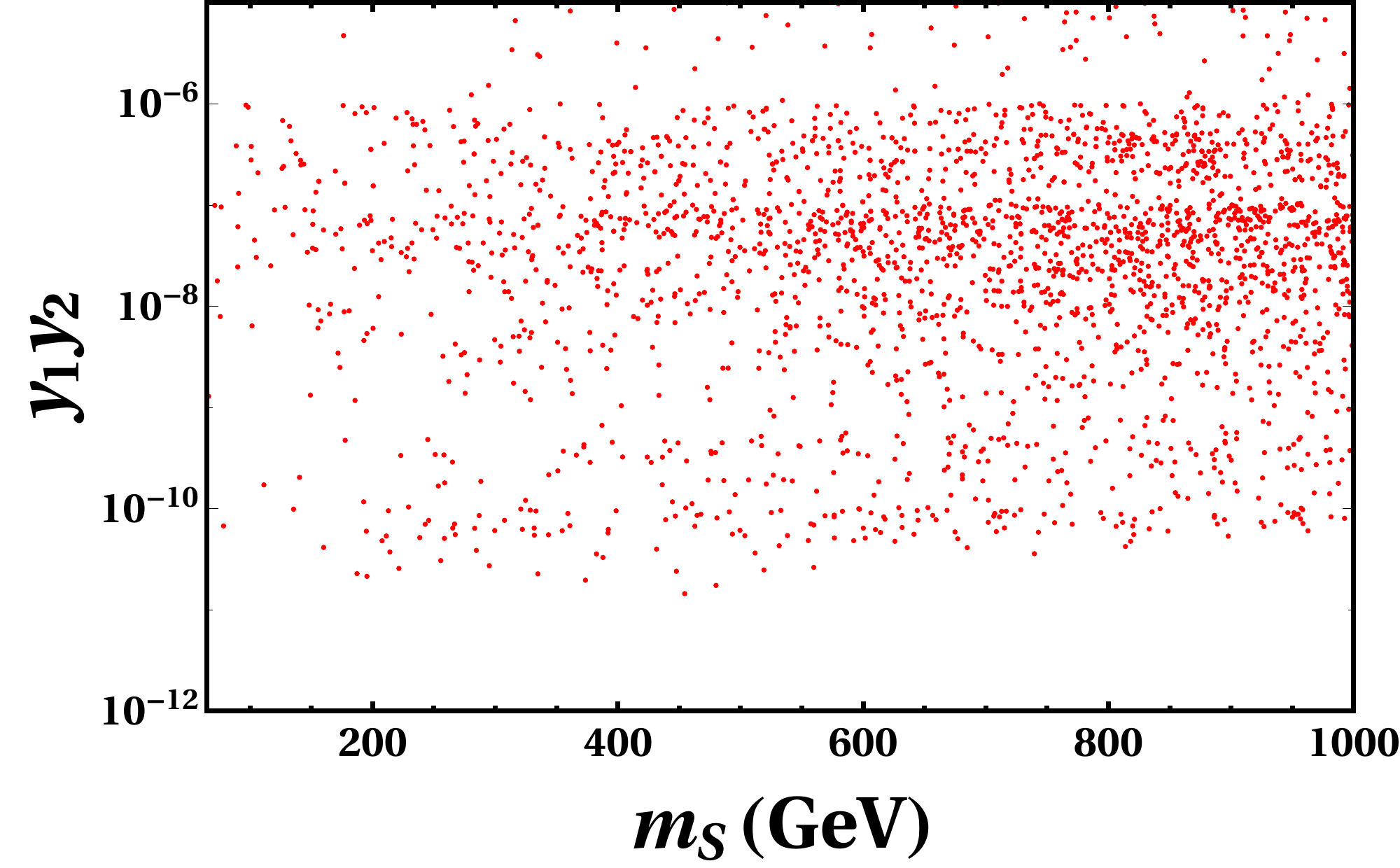}
\caption{Correlation between the relevant coupling and the dark matter mass $m_\psi$  (left), the decaying parent particle mass $m_\chi$ (right) and the associated charged scalar $S^+$ (bottom), satisfying the relic density bound. }
\label{fig:case2_freezein1}
\end{figure}

\begin{figure}[H]
\includegraphics[width=0.48\textwidth]{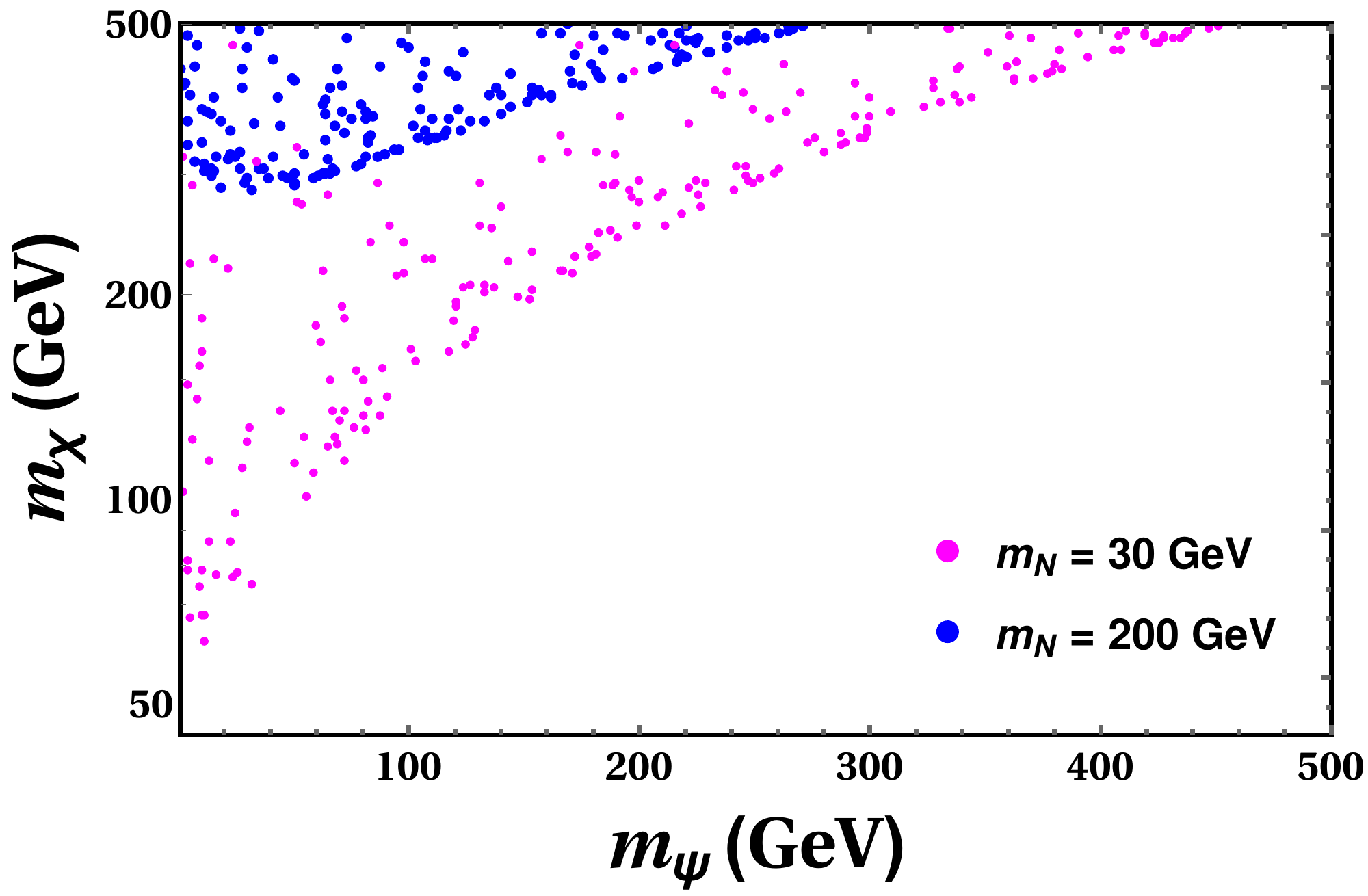}
\includegraphics[width=0.48\textwidth]{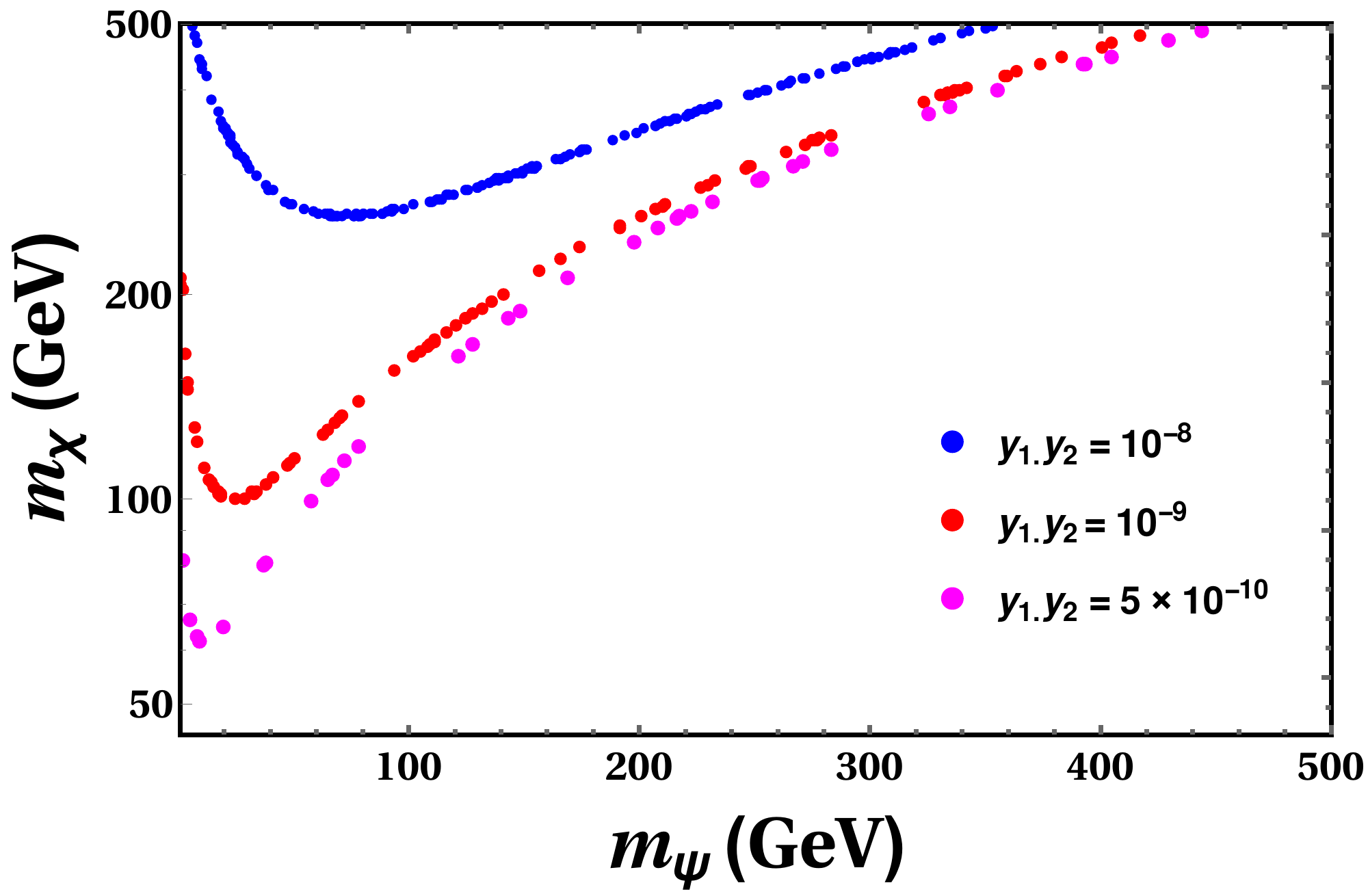}
\caption{Correlation between $m_{\psi}$ and $m_{\chi}$ in the case of freeze-in with 3-body decay of $\chi^{+}$. The {\it left} plot allows $m_S$ and $y_1y_2$ to vary within the specified range, while the {\it right} plot has $m_S=800$ GeV and three different indicative $y_1y_2$ values. $m_N$ is fixed at 30 GeV in the {\it right} plot.}
\label{fig:case2_freezein2}
\end{figure}

In  Fig.~\ref{fig:case2_freezein1} the coupling combination $y_1y_2$ is plotted against $m_\psi$ (top left figure),  $m_\chi$ (ropt right figure) and $m_S$ (bottom figure). Generically, the product of the coupling below $10^{-10}$ is not possible, with a few points between $10^{-11}$ and $10^{-10}$ corresponding to relatively heavier $\chi^{+}$ and lighter $m_S$ so that the propagator factor compensates for the small coupling. For lighter $\chi^{+}$ below 200 GeV, the coupling is mostly larger than $5\times 10^{-10}$, which is filtered down to slightly lower than $10^{-10}$ for masses between 200 and 400 GeV, above which it is possible to go down by one more order. The correlation between $m_\chi$ and $m_\psi$ is plotted in Fig.~\ref{fig:case2_freezein2}, where the effect of $m_N$ is indicated with two illustrative values of $m_N=30$ and 200 GeV. The effect mostly arises from the kinematic limit constraint of $m_\chi> m_\psi+m_N$. In the right-hand-side plot in Fig.~\ref{fig:case2_freezein2}, the effect of the coupling on this correlation is indicated for three different choices of couplings, for a fixed $m_N=30$ GeV. Note that the decay width depends on the inverse powers of its mass, and therefore, require larger values of the couplings for heavier $\chi^{+}$, as clearly indicated in the three cases presented in the plot.

\begin{figure}[H]
\includegraphics[width=0.45\linewidth]{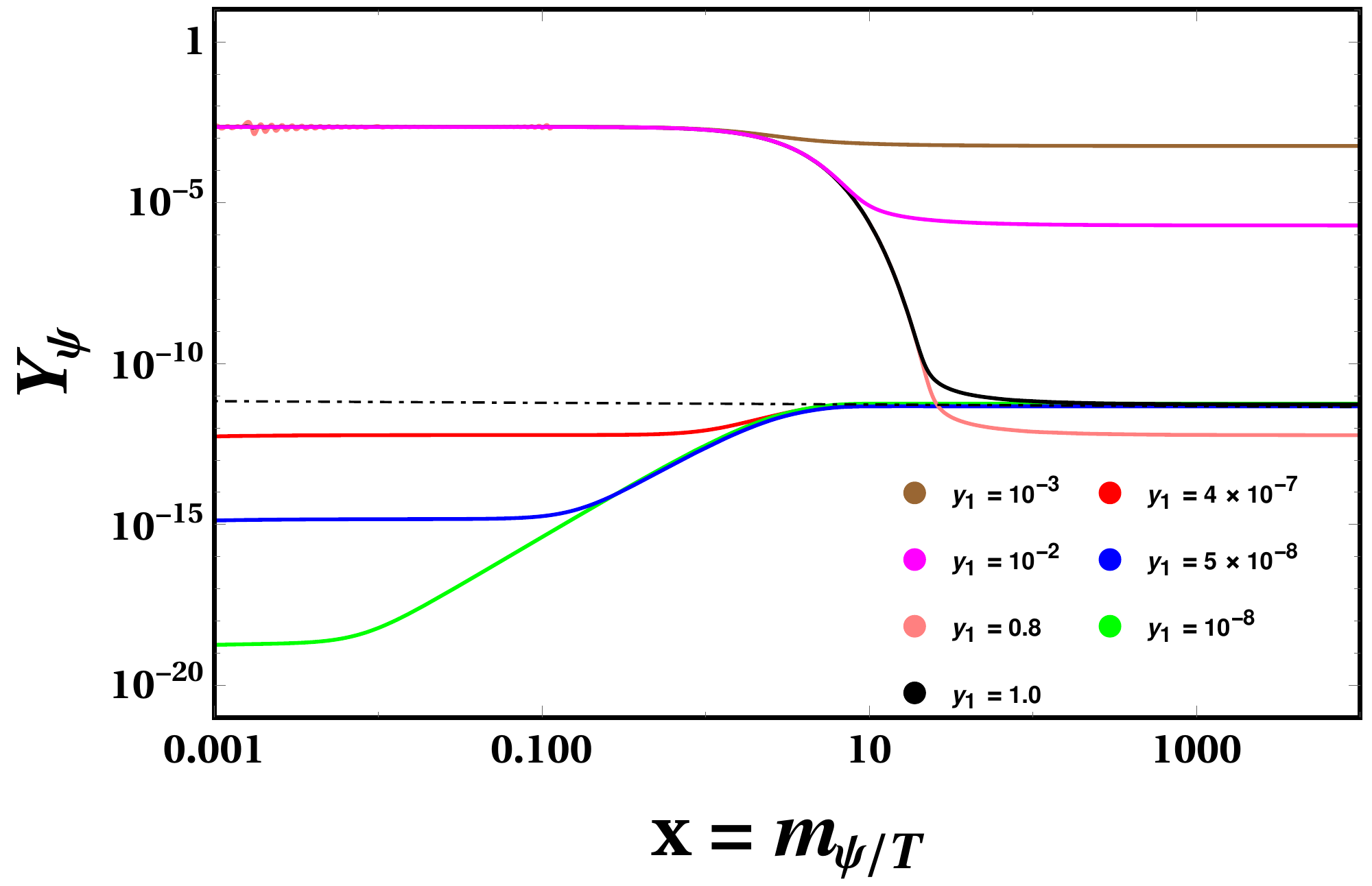} \hskip 5mm
\includegraphics[width=0.45\linewidth]{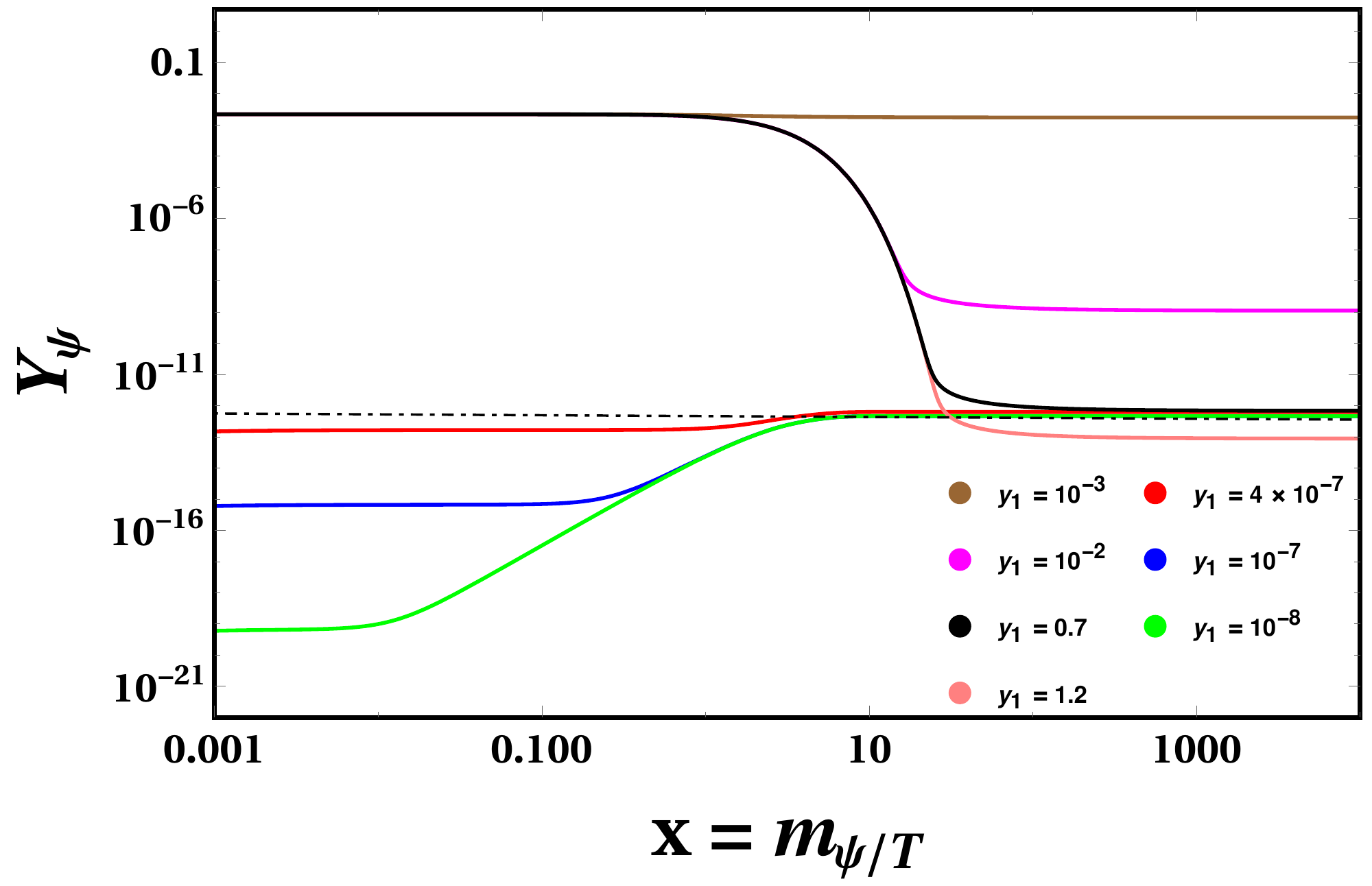}
\caption{{Plots showing interplay of the thermal and non-thermal production of $\psi$ for various values of $y_1$. The non-thermal production of $\psi$ from $\chi$ decay is kept fixed with $y_1y_2=10^{-10}~(\textit{left})$ and $y_1y_2=10^{-6}~(\textit{right})$. The dot-dashed lines represent the correct relic density for the specific benchmark point $m_S$=50 GeV, $m_{\psi}$=80 GeV, $m_{\chi}$=120 GeV in the \textit{left} plot and $m_S$=463 GeV, $m_{\psi}$=941 GeV, $m_{\chi}$=972 GeV in the \textit{right} plot. $m_N$ is fixed at 30 GeV.}}
\label{newplot}
\end{figure}

\noindent\emph{\textbf{In the presence of annihilation:}} The third situation listed above, which includes the effect of sizable annihilation cross sections is interesting, however not usually considered in the literature. In most situations studied, a single coupling parametrised the interactions. Freeze-in mechanism requiring this to be small naturally makes the effects of annihilation irrelevant. However, as we demonstrate here, in the presence of an additional coupling such as present in the scenario discussed in this article provides a natural framework where along with the thermal production, significant amount of DM is produced through the decay of the associated particle. In presence of annihilation, this leads to an interesting interplay of the thermal and non-thermal production of $\psi$. When $y_1$ and $y_1y_2$ both are small, $\psi$ is not in equilibrium with the thermal bath and is only produced from the decay of $\chi$. However, $S^+$, being Gauge produced, if $y_1$ increases, $\psi \psi \to S S $ forward and backward scatterings also produce $\psi$ from the thermal bath and slowly drive it towards equilibrium in the early Universe. This feature is illustrated in Fig. \ref{newplot}, where the yield of $\psi$ is plotted against $m_\psi$ for a fixed $y_1y_2$ but for different $y_1$ values. The green curve can be considered as the typical non-thermal production via freeze-in due to the minuscule $\langle \sigma v \rangle$ (corresponding to $y_1 \sim 10^{-8}$). The blue and the red line correspond to larger $y_1$ which clearly show the increasing effect of the thermal production of $\psi$ from the bath in the early Universe. However, for such values of $\langle \sigma v \rangle$, the non-thermal decay is still the dominant production channel and we see the effect of saturation of yield when it becomes Boltzmann suppressed, similar to the freeze-in scenario. Now, on increasing $y_1$ further, the thermal production becomes more dominant and the typical freeze-out scenario is recovered, which can be seen from the brown, pink, magenta and the black line. Similar to the WIMP freeze-out, these cases give overabundance or underabundance depending on the $\langle \sigma v \rangle$ value considered.

A few benchmark points compatible with this scenario are presented in Table~\ref{table:finfout}\footnote{In Fig. \ref{newplot} and Table. \ref{table:finfout}, large values of $y_1$ is taken to show the effect of the interplay between the thermal and non-thermal production of $\psi$. However, direct detection diagrams in one loop constrains this coupling to $y_1 \lesssim 0.05$. This implies that this coupling in the allowed limit is not sufficient to satisfy the observed relic density, but it can make the thermal production dominant over the non-thermal case and lead to overabundance. We will show later on in the discussion of freezeout that coannihilation and mediator annihilation channels can satisfy the observed relic even with small $y_1$.}, indicating the cases when pair annihilation and co-annihilation channels contribute significantly. For clarity, we have included the respective cases with small $y_1$, where the effect of annihilation is absent, and the correct relic density is not achievable. While the results are presented for a fixed $m_N=30$ GeV, we have checked that it produces the relic density within the same order of magnitude for a range of $m_N$ from 1 GeV to a couple of 100 GeV.
\begin{table}
\centering
\small
\begin{tabular}{c|c|c|c|c|c}
\hline \hline
${\bm m_\chi}$ & $\bm{m_\psi}$ & $\bm{m_S}$ & $\bm{y_1y_2}$ & $\bm{y_1}$ & $\bm{\Omega h^2}$ \\ \hline\hline
\multirow{3}*{515.5} & \multirow{3}*{485.4} & \multirow{3}*{176.3} & \multirow{3}*{$4.7 \times 10^{-6}$} & $10^{-8}$&0.119\\ \cline{5-6}
&&&&$10^{-2}$&$2.1\times 10^{5}$\\ \cline{5-6}
&&&&0.5&0.12\\ \cline{5-6}
&&&&$1.0$&$0.008$\\ \hline
\multirow{3}*{394.5} & \multirow{3}*{352.9} & \multirow{3}*{338.5} & \multirow{3}*{$1.1 \times 10^{-8}$} & $10^{-8}$&0.119\\ \cline{5-6}
&&&&$10^{-2}$&$2.2 \times 10^{6}$\\ \cline{5-6}
&&&&0.98&0.12\\ \cline{5-6}
&&&&$1.5$&$0.02$\\ \hline
\multirow{3}*{460.6}&\multirow{3}*{300.2}&\multirow{3}*{214.1}&\multirow{3}*{$3.72 \times 10^{-11}$}&$10^{-8}$& $0.119$\\ \cline{5-6}
&&&&$10^{-2}$&$5.9 \times 10^{5}$\\ \cline{5-6}
&&&&0.67&0.12\\ \cline{5-6}
&&&&1.0&0.02\\ 
\hline\hline
\end{tabular}
\caption{Benchmark points showing the effect of the addional thermal production of $\psi$ with normal non-thermal production from $\chi^+$ decay. For negligible $y_1$, the correct relic density is achieved by non-thermal production only. Increasing $y_1$ enhances the thermal production of $\psi$ through $S S \leftrightarrow \psi \psi$ channel in the early Universe, suppressing the non-thermal production. Therefore, the freeze-out scenario is recovered, where overabundance or underabundance is dictated by the choice of $y_1$. All masses are in GeV.  The mass of the neutral fermion is set to $m_N=30$ GeV in all cases.}
\label{table:finfout}
\end{table}
It is clear from Fig.~\ref{newplot} that based on the couplings, the DM is either following freeze-in saturation or is driven towards thermal freezeout. This is largely due to the coupling $y_1$ which, for large values, is responsible for sufficient energy exchange between $S^+$ and $\psi$ in the early Universe, therefore, suppressing the non-thermal production of $\psi$. If the production and annihilation channels of $\psi$ could be considered independently, then it would be possible to restrict its production to only non-thermal decay of $\chi$ and annihilate through some other coupling which does not contribute to the energy exchange of DM with thermal bath. A possible way-out is to consider cases like mediator driven annihilation and conversion as the possible modes of annihilation of $\psi$. It is well known that for very small mass splitting between $\psi$ and  $\chi$, $\chi \chi \to$ SM SM processes contribute substantially to the relic density of $\psi$. In this model, these processes are Gauge mediated, hence independent of $y_1$. Hence, even if we take both $y_1$ and $y_2$ small enough to ensure only non-thermal decay of $\chi^+$, the production and the annihilation of $\psi$ still remains independent of each other. Ref.~\cite{Junius:2019dci} discusses a similar situation and on solving out-of-equilibrium Boltzmann eqns.~(see Eqn.~9 and 10 of Ref.~\cite{Junius:2019dci}) for DM candidate and the mediator, an interesting situation is obtained which can be interpreted as an intermediate stage between freeze-in and freeze-out scenario, where the non-thermal over production can be tamed by sufficient annihilation. However, as seen from the parameter choice of this plot, this is valid only for a small mass splitting between the DM candidate and the mediator, which is not of much importance for our discussion since the viable parameter space is very much contrived.

%
\section{Exploring freeze-out possibility}
\label{Freezeout}
The model can also explain thermal dark matter, such as the standard WIMP scenario through thermal freeze-out, which is viable over a large regions of parameter space. In this case, it is assumed that the DM candidate $\psi$ is already in equilibrium with the thermal bath. Hence we no longer require the DM interaction coupling $y_1$ to be very small unlike freeze-in scenario, therefore it is considered to be of the order of unity. All the possible annihilation diagrams are given in Fig.~\ref{fd4}. Based on the mass splitting between the dark sector particles ($\chi^+$ and $\psi$), the relic density will be dominated by either annihilation of DM into SM particles (Region 1) or co-annihilation between the dark fermions (Region 2). Fig.~\ref{fd4}(a) represents the only diagram possible when $\psi$ pair annihilates into $S^{+}S^{-}$ pairs through $t$-channel mediation of $\chi^{+}$. Fig.~\ref{fd4}(b)-(d) represent the co-annihilation channels which become effective only when the mass splitting between incoming particles is very small. {The model is implemented in Feynrules~\cite{Alloul:2013bka} and the freeze-out relic density computation is done using micrOMEGAs~\cite{Belanger:2014vza}.}

The parameter space is classified into two regions of $m_{\chi} \gg m_{\psi}$ and  $m_{\chi} \sim m_{\psi}$, with the co-annihilation channels becoming important in the latter case. We shall discuss these two regions separately in the following subsections.

\subsection{Region with large mass splitting: $m_{\chi} \gg m_{\psi}$ }
The only pair annihilation of the dark matter in this region proceeds through the $t$-channel to a final state of $S^{+}S^{-}$ pairs shown in Fig.~\ref{fd4}(a), with the  cross section proportional to $y_1^4$. The co-annihilation process have negligible contribution as the cross section for such process goes like the exponential of mass splitting between the dark matter and the partner, $e^{-(m_\chi-m_\psi)/m_\psi}$~\cite{Griest:1990kh}. To study the relic compatible parameter space regions, we scan over the available/relevant range of parameters and compute the relic density using micrOMEGAs. As in the earlier case, the irrelevant quartic couplings, $\lambda$ and $\lambda_1$ are fixed at 0.01, and the Yukawa coupling dictating the neutrino mass are set at $y_{N_i} \approx 10^{-8}$.  The mass of the neutral fermions are set to $m_N=30$~GeV, while the Yukawa coupling between $S^{+}N\ell$ is taken as $y_2=10^{-6}$.  The parameters, which are relevant to the annihilation process here are varied randomly in the range given in Table~\ref{table:freezeout1}.

\begin{table}[H]
\centering
\begin{tabular}{c |c }
\hline \hline
1 GeV $ \le m_{\psi} \le $ 1 TeV  &     $m_{\chi} = m_{\psi} + (100, ~800)$ GeV\\ \hline
 65 GeV $\le m_S \le  $ 1 TeV & $0.0 \le y_1 \le 3.0$\\ \hline \hline
\end{tabular}
\caption{Range of parameters considered for the scan in Region 1: $m_{\chi} \gg m_{\psi}$ with thermal freeze-out of dark matter $\psi$.}
\label{table:freezeout1}
\end{table}

\begin{figure}[H]
\centering
\includegraphics[scale=0.5]{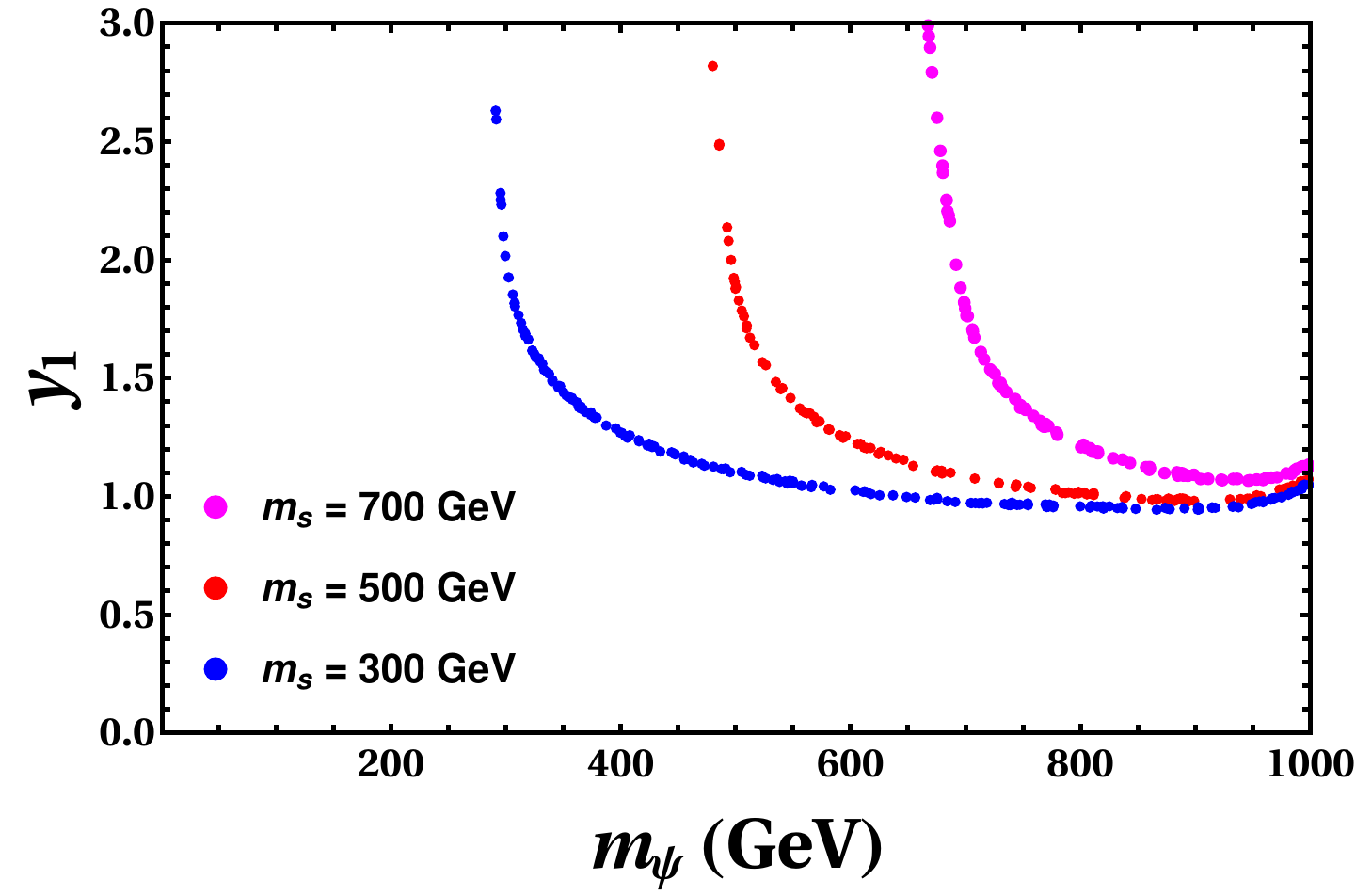}~~
\includegraphics[scale=0.5]{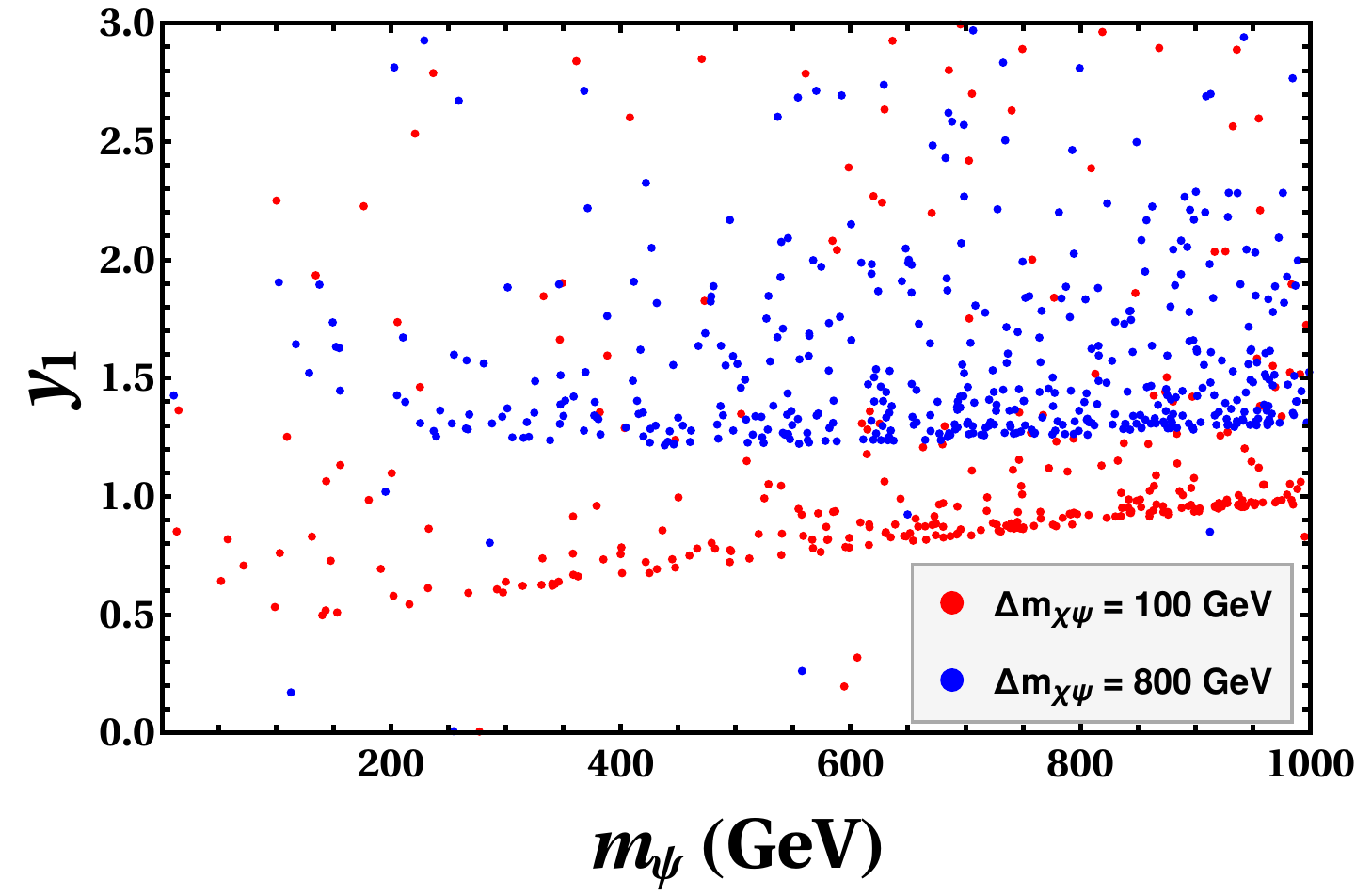}
\caption{$y_1$ vs.~$m_{\psi}$ for specific values of $m_S$ in agreement with the measured DM relic density; Left: for a fixed propagator mass of $m_\chi=1$ TeV; Right: for varying $m_\chi$ with the mass splitting fixed at $m_{\chi}-m_{\psi}=100$~GeV (red) and 800~GeV (blue) keeping $65\le m_S\le 1000$ GeV.}
\label{ann_frzout}
\end{figure}

In Fig.~\ref{ann_frzout}, the variation of the coupling $y_1$ is plotted against the DM mass. The $\psi \psi \rightarrow S^{+} S^{-}$ annihilation channel opens up at different $S^{+}$ values, as shown in the left plot in Fig.~\ref{ann_frzout}, for a fixed $\chi^{+}$ mass of 1 TeV. 
In the right plot of Fig.~\ref{ann_frzout}, $m_{\chi}$  is varied for the entire range of $m_{\psi}$ keeping the mass splitting fixed at two specific values. For smaller mass splitting, a relatively lighter propagator reduces the demand on the requirement of large coupling. However, to be compatible with the direct detection limits of $y_1\le 0.05$, highly degenerate region of the parameter space may be required, as explained towards the end of the next subsection. It is to be noted here that if $S^+$ was not charged, the one loop penguin diagrams (Fig. \ref{fig:dd}) contributing to direct search would not contribute, but still the strength of the new physics coupling would determine the interplay between the nonthermal and thermal production, even leading to freezeout for a wider parameter space. Because of this generic feature, we have kept the full range of $y_1$ in the analysis, although  for our particular case freezeout is restricted to a narrow window of the parameter space.

\subsection{Region with almost degenerate case: $m_{\chi} \sim m_{\psi}$}
In this region, as the mass difference between $\chi^{+}$ and $\psi$ is very small, the effect of the co-annihilation channels become important. Unlike Fig.~\ref{fd4}(a), these channels provide correct relic density even when $m_{\psi} \leq m_S$. This is because $\psi$ can now annihilate into $S^{+}$ and a photon or a $Z$ boson provided $2m_{\psi} \geq m_S/(m_S+m_Z)$ (Fig.~\ref{fd4}(c),(d)). Setting a non-zero value of $\lambda_1$ enables the process considered in the diagram in Fig.~\ref{fd4}(b), although the effect of this cross section is very small. However, since $\lambda_1$ is fixed at a small value (= 0.01) in the analysis, this contribution is always neglible. We consider the ranges of parameters given in Table~\ref{table:region2} while performing the scan.

\begin{table}[H]
\centering
\begin{tabular}{c |c }
\hline
\hline
1 GeV $ \le m_{\psi} \le $ 1 TeV  &     $m_{\chi}=  m_{\psi}$ +1, 5, 10 GeV\\ \hline
 65 GeV $\le m_S \le  $ 1 TeV & $0.0 \le y_1 \le 3.0$\\ \hline\hline
\end{tabular}
\caption{Range of parameters considered for the scan in Region 2: $m_{\chi} \sim m_{\psi}$ with thermal freeze-out of dark matter $\psi$.}
\label{table:region2}
\end{table}

\begin{figure}[H]
\includegraphics[width=0.48\textwidth]{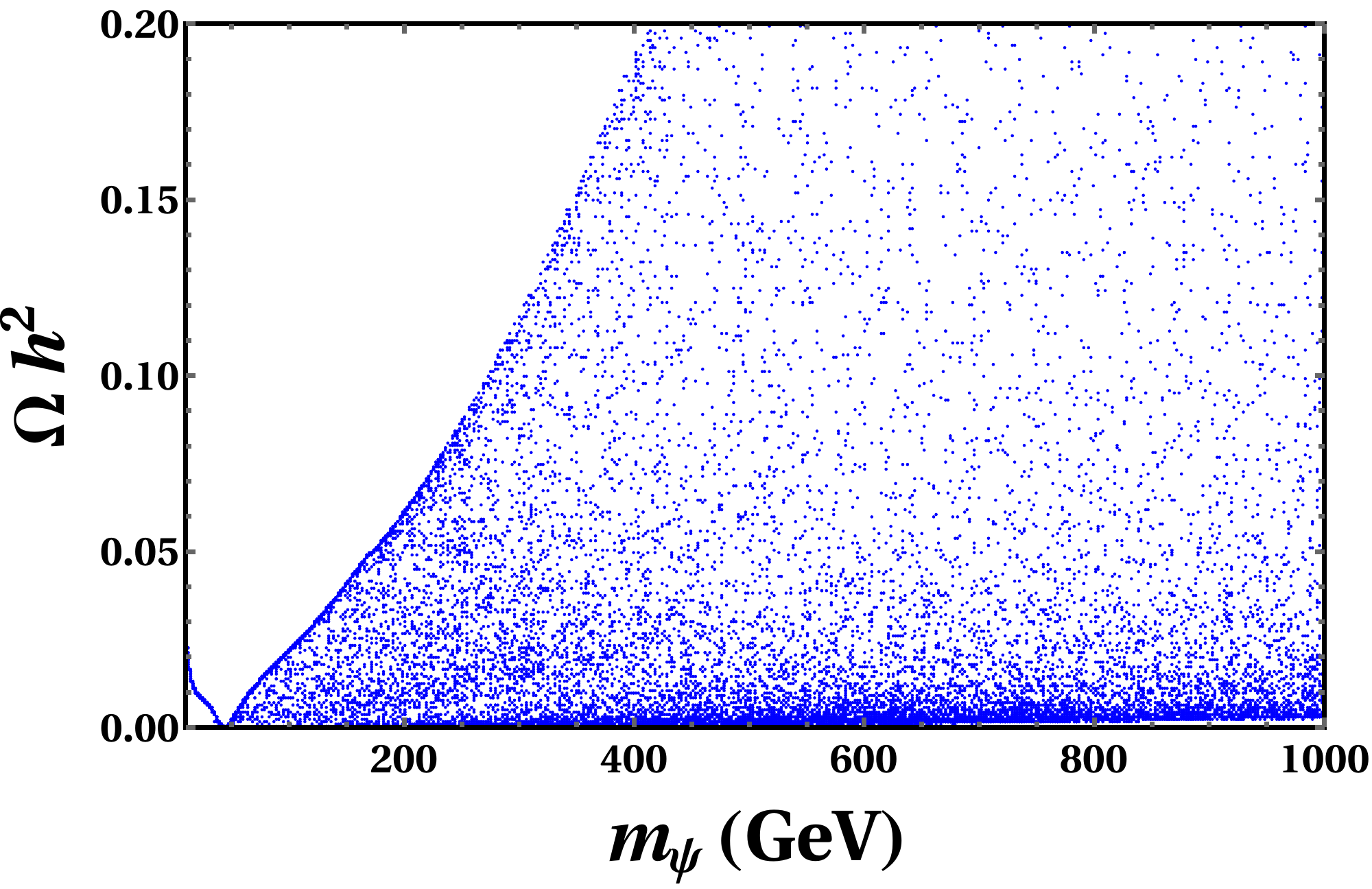}
\includegraphics[width=0.48\textwidth]{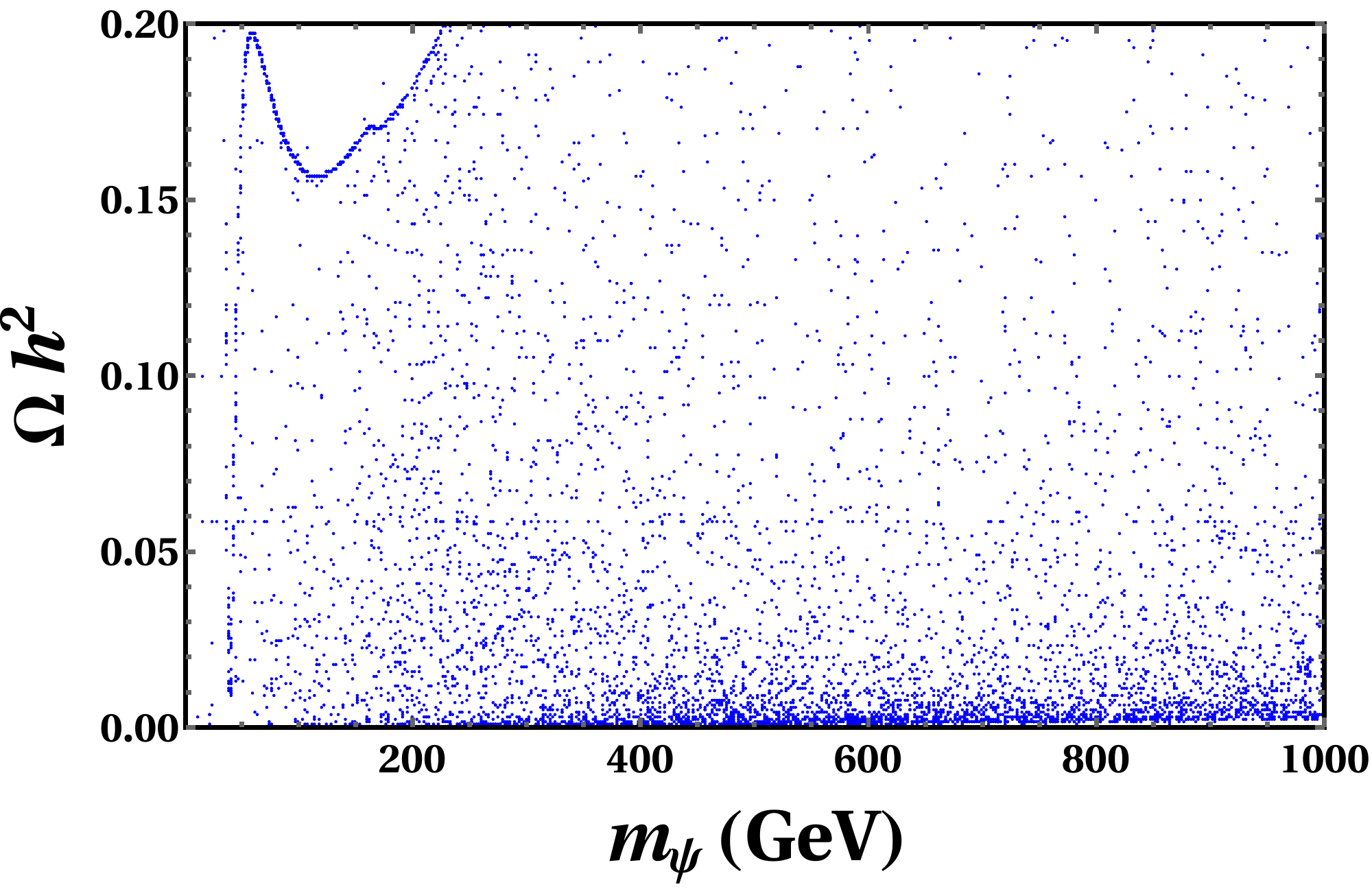}
\caption{DM relic density vs.~DM mass for mass splitting between dark sector particles fixed at 1 GeV (left) and 5 GeV (right). For larger mass splitting (right), the effect of $Z$-mediated $s$-channel cross sections arising due to $\chi^{+}$ annihilation become negligible, and co-annihilation channels dominate increasing the relic density in the low DM mass regime within observed limits.}
\label{coann_1}
\end{figure}
In Fig.~\ref{coann_1}, the variation of DM relic density is plotted with DM mass for two different mass splittings between $\chi^{+}$ and $\psi$. In the left panel, the splitting is fixed at 1 GeV. For such a small splitting, the annihilation of the other dark sector particle $\chi^{+}$ into SM fermions also contributes to the relic density along with the diagrams given in Fig.~\ref{fd4}. All these additional channels being $Z$-mediated, we see the $Z$-resonance at $m_{\psi}$=45 GeV. After the resonance, these cross sections decrease with increase in $m_{\psi}$ making the co-annihilation channels important. But if the mass splitting is made larger (5 GeV as shown in the right plot), then the effect of mediator annihilation, ie, the contribution of these $Z$-mediated diagrams becomes negligible compared to $\psi$ co-annihilation and pair annihilation, so that even the low mass regime of $m_{\psi}$ supports the correct relic density.
\begin{figure}[H]
\centering
\includegraphics[width=0.55\textwidth]{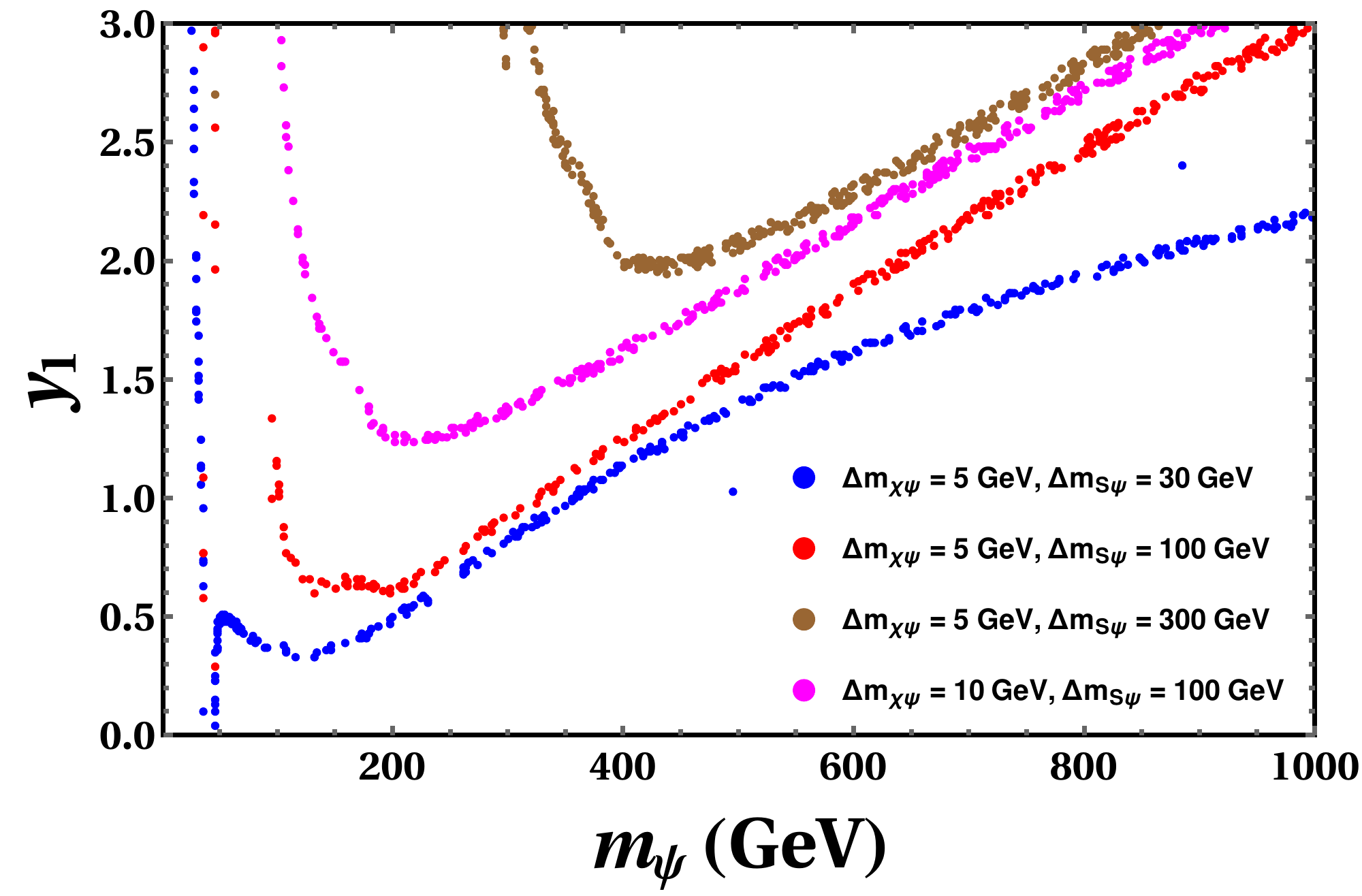}
\caption{$y_1$ vs.~$m_{\psi}$ plot indicating the significance of the co-annihilation channels, with the mass splitting $\Delta m_{\chi\psi}=m_\chi-m_\psi=5$ and 10 GeV, and for specific values of $\Delta m_{S\psi}=m_S-m_\psi$.}
\label{coann_2}
\end{figure}
In Fig.~\ref{coann_2}, the variation of $y_1$ vs $m_{\psi}$ is plotted with the mass splitting between $\psi$ and $\chi^{+}$ ($\Delta m_{\chi\psi}$) and the splitting between $\chi^{+}$ and $S^{+}$ ($\Delta m_{S\psi}$)  fixed at specific values. The dip at $m_\psi=45$~GeV indicates a broad $s$-channel $Z$-resonance arising due to $\chi^+ \chi^-$ mediator annihilation into SM fermions. The corresponding cross section being proportional to $e^{-2\Delta m_{\chi\psi}/m_\psi}$, it becomes irrelevant at mass splitting of 10~GeV, as indicated by the absence of such a dip in the deep pink curve. The co-annihilation channels, on the other hand, has a slightly more complex dependence on the masses. Considering the Feynman diagrams given in Fig.~\ref{fd4}, there are three channels that could contribute. Two $s$-channel processes mediated by $S^{+}$ and a $t$-channel process mediated by $\chi^{+}$. Both the $s$-channel processes have $S^{+}$ in the final state as well. This brings in a tug of war between the resonant condition, $m_\chi+m_\psi=m_S$ and the favourable phase space for the lighter final state particles, consequently disfavouring the $\chi^{+}\psi\rightarrow S^{+}h$ channel. On the other hand, the other $s$-channel process with $S^{+}\gamma$ in the final state becomes significant around $m_S=m_\chi+m_\psi\sim 2m_\psi$. The $t$-channel process takes over at larger $m_\psi$, and the cross section slowly saturates. This is indicated in the plots at higher $\psi$ masses. Heavier $S^{+}$ requires larger couplings to compensate for the diminishing effects of the propagator and narrowing phase space. This is clearly visible in the plots with larger $\Delta m_{S\psi}$ values.
As soon as $m_{\psi}$ becomes larger than $m_S$, the  pair annihilation of $\psi$ becomes dominant and the correct relic density is observed for $m_{\psi}$ up to the TeV scale. 
Summing up, the solutions compatible with the required $y_1 \lesssim 10^{-5}$ is possible with highly degenerate regions with $\Delta m_{\psi\chi} \lesssim 5$ GeV in a narrow mass range of $m_\psi\sim\frac{m_Z}{2}$.

%
\section{Collider signatures}
\label{Collider_discussion}
\begin{figure}[H]
\begin{center}
\begin{subfigure}{0.48 \textwidth}
\includegraphics[width=\textwidth]{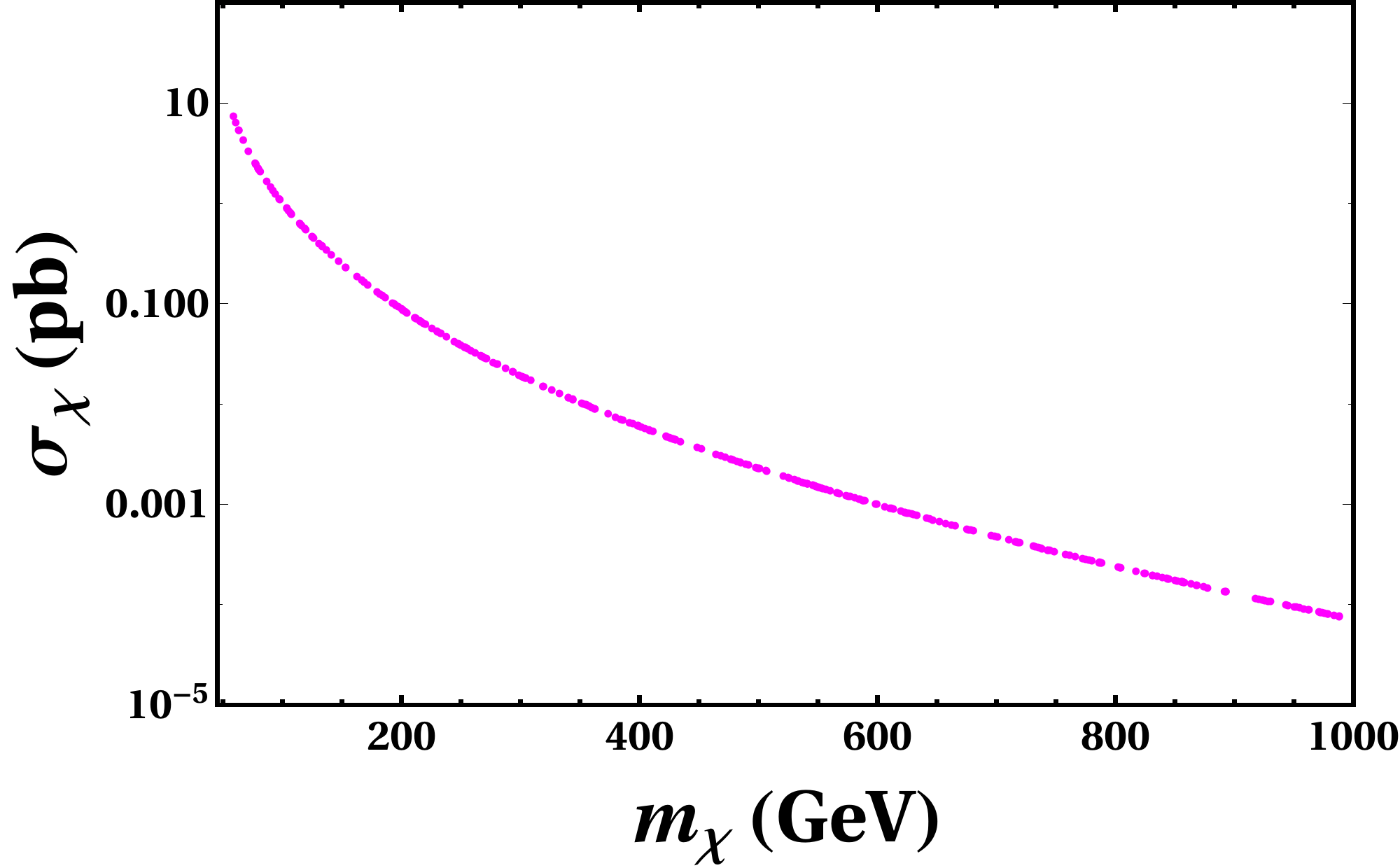}
\caption{ }
\label{fig_chi_cs}
\end{subfigure}
\hfill
\begin{subfigure}{0.48 \textwidth}
\includegraphics[width=\textwidth]{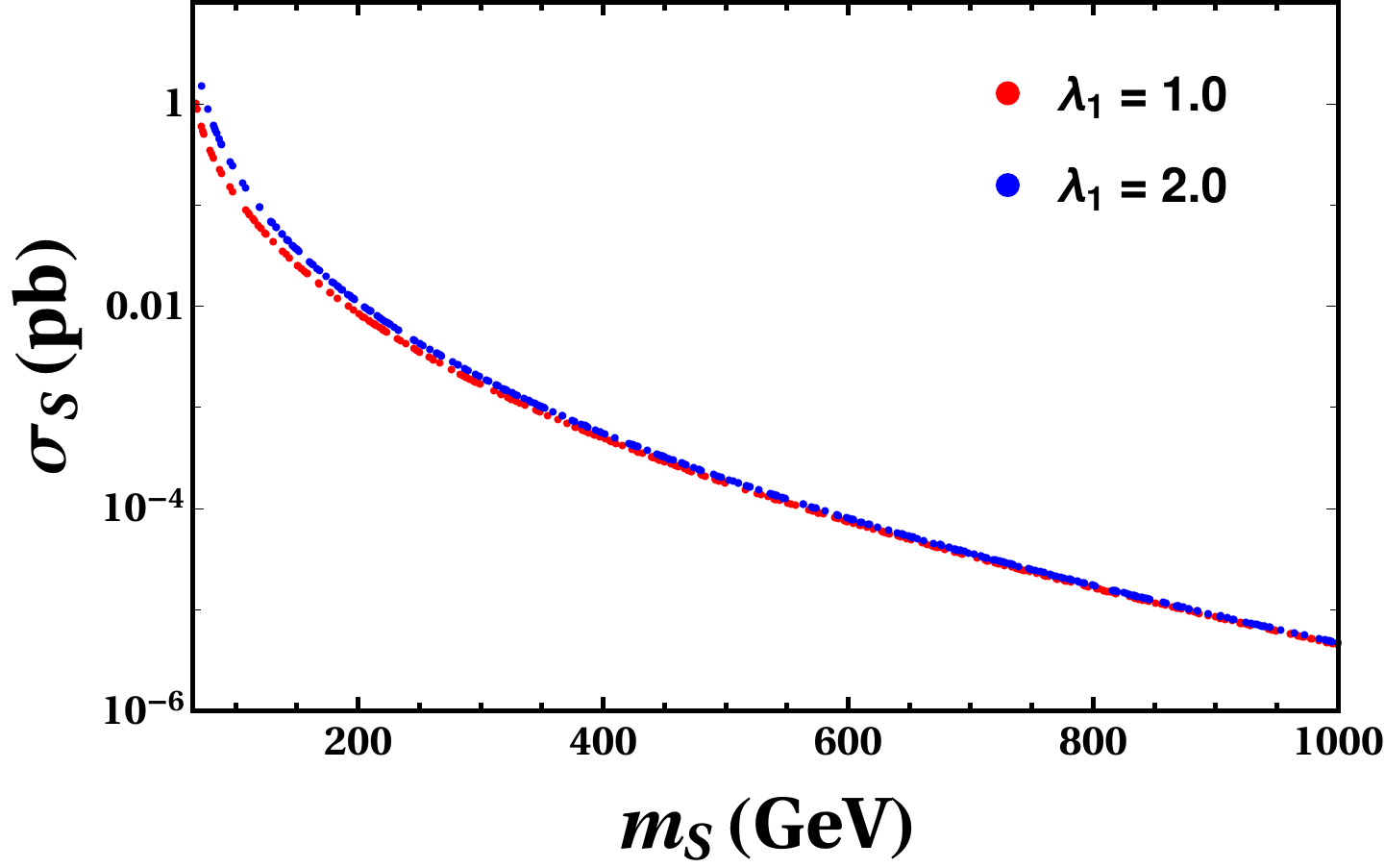}
\caption{ }
\label{fig_SS_cs}
\end{subfigure}
\caption{Production cross section of $\chi^{+}\chi^{-}$ (a) and $S^{+}S^{-}$  (b) pair at 14 TeV LHC for different values $m_\chi$ and  $m_S$, respectively.}
\end{center}
\end{figure}

The signatures of the model can be traced in colliders like the LHC through the gauge production and subsequent decay of the charged partner fermion, $\chi^{+}$ and the charged scalar $S^{+}$.  We plot the cross section for pair production of $\chi^{+}$ ($\sigma_{\chi}$) at the 14~TeV LHC in Fig.~\ref{fig_chi_cs}. At $m_\chi=200$ GeV, the cross section is of the order of 100 fb, which falls down to 0.5~fb for $m_\chi=1$~TeV.  The charged scalar $S^+$, on the other hand, has much smaller cross section ($\sigma_S$) of about 1~fb for $m_S=100$ GeV, which reduces to about 0.1~fb for $m_S=500$~GeV as seen from Fig.~\ref{fig_SS_cs}. Subprocesses contributing to $\sigma_{\chi}$ and $\sigma_S$ are $p p \to \chi^+\chi^-/S^+S^-$ via $Z/\gamma$ s-channel diagrams. The decays of $\chi^{+}$ and $S^{+}$ are controlled by the Yukawa couplings $y_1$ and $y_2$. When both the couplings are large enough, the decays are instantaneous and happens at the interaction point itself where the particles are produced.  On the other hand, for small values of the couplings  the decay width could be considerably smaller so that the decays happen away from the interaction point, leaving distinct signatures characteristic to such Long-Lived Particles (LLP). Indeed, such LLP arising in different contexts are actively studied for their signatures at LHC and other exclusively designed colliders like MATHUSLA in the past few years. Signatures typically include {\em (i) Displaced Vertex}, where the LLP decay within the detector, but away from the interaction point; {\em (ii) Heavy Stable Charged Particle (HSCP) tracks}, where charged LLP decaying outside the detector, leaving charged tracks similar to that of the muon, but with distinct features arising from slower speed and larger energy deposition along its flight through the materials of the detector or {\em (iii) Disappearing Tracks}, where an isolated track stopping before hitting the outer layers of the silicon tracker and without any associated energy deposition in the calorimeters or muon chamber. Experimental searches by CMS~\cite{Sirunyan:2018vlw,Sirunyan:2018njd, Sirunyan:2018pwn}, ATLAS~\cite{Aaboud:2019opc, Aaboud:2018kbe,Aaboud:2018aqj,Aaboud:2018arf} and LHCb~\cite{Aaij:2017mic,Aaij:2016xmb,Liu:2018wte} considered long-lived neutral scalar decaying in the calorimeter into pair of jets, looking for displaced vertex with negative results leading to constraints on the masses of the corresponding LLP. Considering a gauge-mediated supersymmetry breaking scenario, CMS~\cite{Sirunyan:2019gut} considered non-prompt jets arising from the decay of gluino to gluons and gravitinos, where the gluon jet emerges from a displaced vertex within the tracker. HSCP search by  ATLAS~\cite{ Aaboud:2019trc,Aaboud:2018hdl} and CMS~\cite{Khachatryan:2016sfv} looking for possible candidates in the supersymmetric scenarios have obtained mass limits on the so-called $R$-hadrons (long-lived gluinos or squarks), staus and charginos.  Long time of flight along with presence of anomalously high energy deposits in the silicon detector is basically considered as the signature of the HSCP. Reference~\cite{Aaboud:2018hdl} has used the ionisation information from the pixel subsystems of ATLAS detector to probe the HSCP modelled after $R$-hadron. CMS~\cite{Sirunyan:2018ldc,CMS:2019twi} and ATLAS~\cite{Aaboud:2017mpt} considered disappearing tracks arising from chargino decay to neutralinos and a low-momentum pion, where the neutralino is nearly mass degenerate with the chargino, resulting in its long decay time.  A recent report on the strategies that may be adopted in searching for LLP at colliders like the LHC  discussing simplified models, experimental signatures, details of the backgrounds and possible requirements of detector upgrades presented in Ref.~\cite{Alimena:2019zri}. A comprehensive analysis of LLP arising in FIMP like scenario is presented in Ref.~\cite{Belanger:2018sti},  where the authors have described possible signatures of a long-lived charged fermion decaying into the SM charged lepton and scalar DM candidate. Considering fermionic dark matter as a mixture of $SU(2)_L$ singlet and doublet LLP signatures at the LHC are studied in Ref. \cite{Calibbi:2018fqf}.  Charged dark doublet scalar decaying into fermionic FIMP and charged leptons are considered in \cite{Hessler:2016kwm}.  In addition to the possible searchers at the LHC, there are proposals for dedicated detectors like MATHUSLA \cite{Curtin:2018mvb,Lubatti:2019vkf}, which can potentially look for long-lived neutral particles decaying to two jets or charged leptons outside the LHC detectors. These surface detectors are located so as to look for long-lived neutral particles originating from the interaction point, decaying in its volume into two identifiable jets or leptons.

Many of the signatures disccussed above would be present in the scenario discussed in this work. In this case, the charged dark fermion, $\chi^+$, the $Z_2$-even charged scalar, $S^+$ and the heavy neutrino $N$ are all possible candidates of LLP. The specific signatures of the LLP will depend on many aspects. A particular feature pertaining to the model presented in our case is the possibility of {\em multiple LLP signatures} arising through the cascade decay of the kind $ LLP1 \rightarrow LLP2+X $. In the present case, this is possible whenever the 2-body decay of $\chi^{+}$ is allowed. In such case, $\chi^{+}$ decays to an on-shell $S^{+}$, which itself could be an LLP when its decay is slow, enabled by small $y_2$. However, for large enough $y_2$, $S^{+}$ decays quickly enough without leaving any displaced vertex. In that case, the LLP signature of $\chi^{+}$ is the standard case of usual single LLP. Interesting signature of multiple LLP processes are not investigated in detail in the literature. To illustrate this we shall take a specific case with $c\tau(\chi)=\frac{c}{\Gamma_\chi}\le 50$ cm  and $c\tau(S^{+})\sim 1-10$ m, with the first displaced  vertex falling in the Inner Detector, followed by the second one (that of $S^{+}$ decay) falling in the Calorimeter or Muon Spectrometer. The first displaced vertex will leave a kink on the track, with the accompanying $\psi$ invisible. Distinguishability of this kink depends on the masses and energy of the particles, apart from the granularity of the detectors. While we recognise the fact that such possibilities are complex to probe, with precision of a few microns on position in the pixel detector, and high precision calorimeters it is worth exploring the possibilities. In addition to charged $\chi^{+}$ and $S^{+}$, we have the presence of heavy neutral fermions that couple to the charged leptons through the standard Higgs boson. The Yukawa coupling corresponding to this interaction is small, making $N$ almost always stable within the detector. However, for $N$ heavier than 125~GeV, Yukawa couplings of the order of $10^{-8}$ would lead to its decay within the detector.

We shall consider some preliminary numerical analysis of the above long-lived situations with $\chi^{+}$, $S^{+}$ and $N$ separately in the following subsections.

\subsection{Long-lived charged fermion, $\chi^{+}$}
\begin{figure}[H]
\begin{center}
\begin{subfigure}{0.48 \textwidth}
\includegraphics[width=\textwidth]{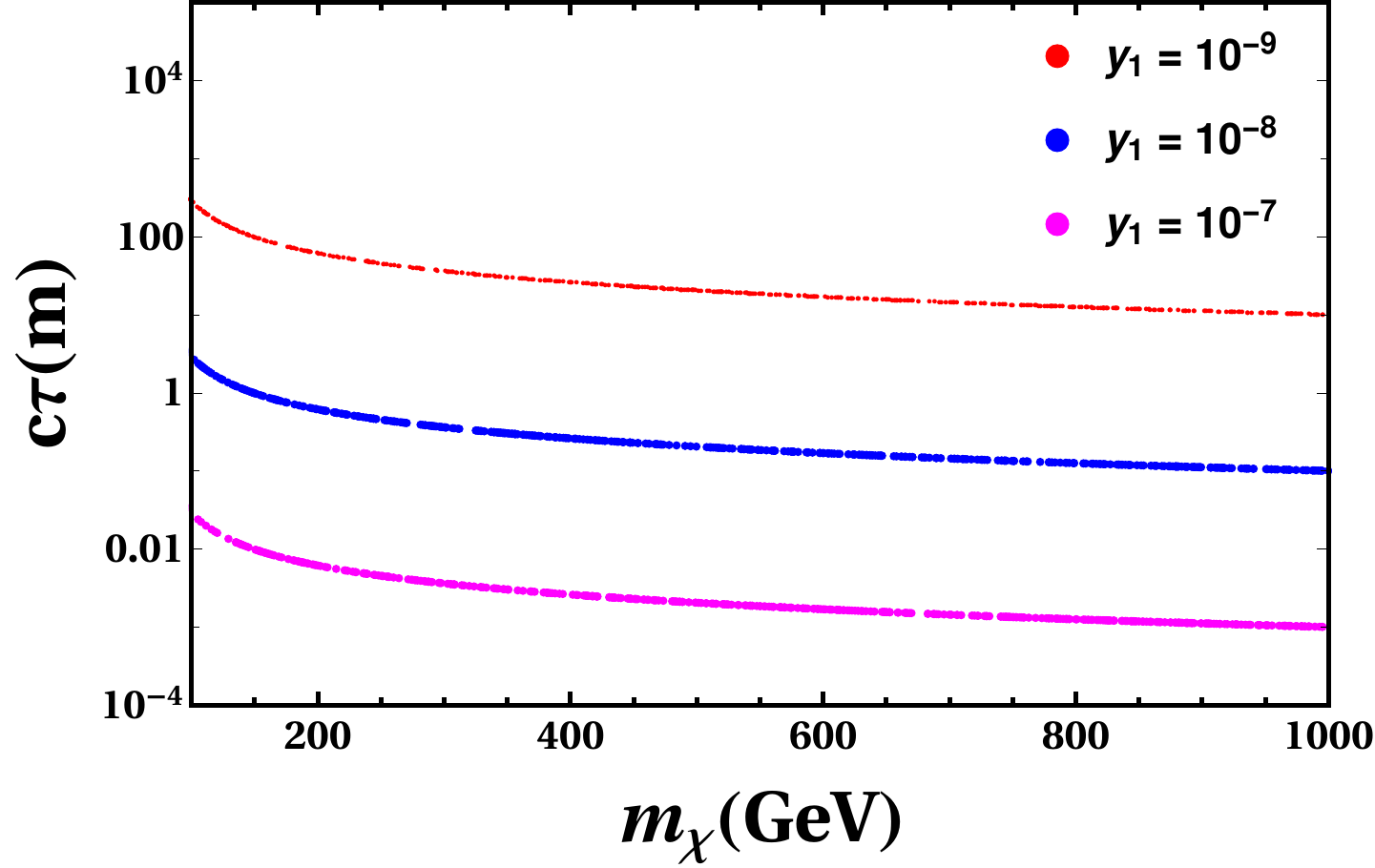}
\caption{ }
\label{fig_chi_ct}
\end{subfigure}
\hfill
\begin{subfigure}{0.48 \textwidth}
\includegraphics[width=\textwidth]{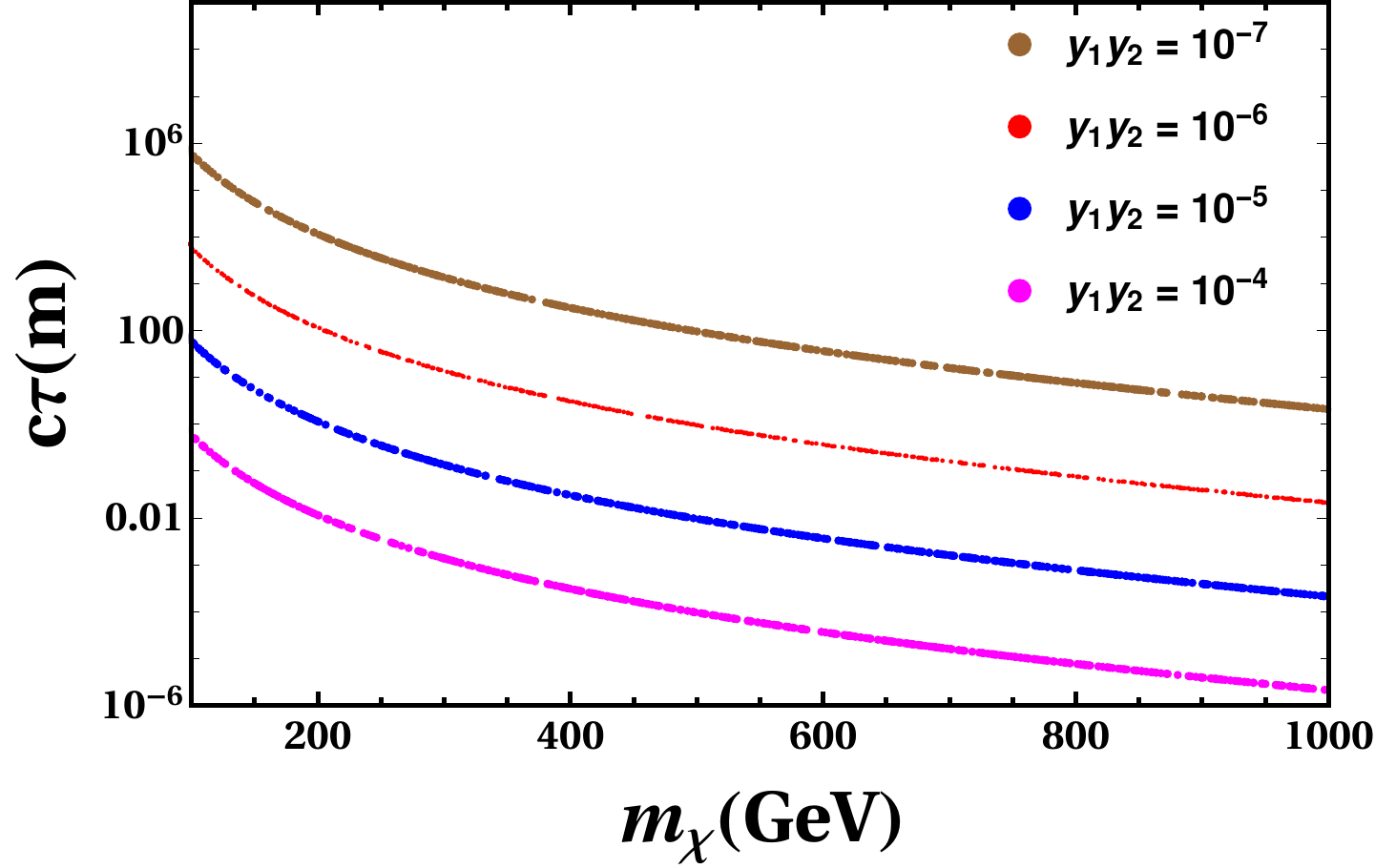}
\caption{ }
\label{fig_SS_ct}
\end{subfigure}
\caption{Decay length vs.~mass of $\chi^{+}$ for different couplings corresponding to $(a)$ 2-body and $(b)$ 3-body decay of $\chi^{+}$.   The neutral particle masses are $m_\psi=1$ GeV and $m_N=30$ GeV, while the charged scalar masses $m_S$ are 65 GeV for case $(a)$ and 1 TeV for case $(b)$.}
\end{center}
\end{figure}

The conditions on $\chi^{+}$ arising from the dark matter considerations depend on its mass and nature of decay. In the kinematic regions where $m_\chi>m_S+m_\psi$, the decay of $\chi^{+}$ is dictated by the Yukawa coupling, $y_1$, the Feynman diagram for which is given in  Fig.~\ref{fd2}. From Fig.~\ref{case1_freezein} it is clear that $\chi^{+}$ lighter than about 200 GeV is not favoured in this case. The compatible couplings are  restricted to $10^{-13}\le y_1\le 10^{-8}$, leading to $c\tau$ in the range of interest of LHC for the larger limits of the coupling. Fig.~\ref{fig_chi_ct} gives the decay lengths for a range of $m_\chi$ and typical values of the coupling, for fixed masses of $m_\psi=1$ GeV and $m_S=65$ GeV.  We have checked the sensitivity with varying $m_\psi$ and $m_S$, and found that the decay length remains in the same order of magnitude for a large range of these masses. Keeping all other parameters the same, the dependence of $c\tau$ on $m_\chi$ is minimal, allowing a wide range of the mass values to be accessible as LLPs at the LHC.  We tabulate the decay lengths for typical values of $m_\chi$ fixing  $y_1=10^{-8}$ in Table~\ref{table_chi2body}. Note that the decay width is proportional to $y_1^2$ in this case, and therefore, $c\tau$ for a specific $y_1$ value can be obtained from what is quoted in table by using the scaling factor ${10^{-16}}/{y_1^2}$.  For example, by choosing a slightly larger $y_1$ value the decay length can be brought down to the centimetre range, which is relevant to the Inner Detector of LHC experiments. At the same time, couplings smaller than $10^{-8}$ leads to $\chi^{+}$ escaping the detector without decaying, leaving charged track in all sections of the detector, very similar to that of the $\mu$ tracks, but with larger energy deposition owing to it being heavier~\cite{CidVidal:2018eel,Belanger:2018sti}. In addition, its time of flight will be larger than that of muon as it would be slower than muons.

\begin{table}[h]
\centering
\begin{subtable}{0.45 \textwidth}
\begin{tabular}{c|c|c|c|c}
\hline \hline
${\bm{y_1}}$ & ${\bm{m_\chi}}$ & ${\bm{m_\psi}}$ & ${\bm{m_S}}$ & ${\bm{c\tau}}$ {\bf (m)} \\ \hline\hline
\multirow{6}*{$10^{-8}$}
&&130&65&55\\\cline{3-5}
&200&65&65&72\\ \cline{3-5}
&&65&130&110\\ \cline{2-5}
&&900&65&63\\\cline{3-5}
&1000&65&65&88\\ \cline{3-5}
&&65&900&710\\  \hline \hline
\end{tabular}
\caption{}
\label{table:LLP2bodydecay}
\end{subtable}
\begin{subtable}{0.45 \textwidth}
\begin{tabular}{c|c|c|c|c}
\hline \hline
${\bm{y_1y_2}}$ & ${\bm {m_\chi}}$ & ${\bm{m_\psi}}$ & ${\bm{m_S}}$ & ${\bm{c\tau}}$ {\bf (m)} \\ \hline\hline
\multirow{7}*{$10^{-6}$}&40&1&65&166\\\cline{2-5}
&100&60&120&69\\ \cline{2-5}
&\multirow{2}*{500}&400&300&1\\ \cline{3-5}
&&400&1000&155\\\cline{2-5}
&\multirow{3}*{1000}&800&800&1\\ \cline{3-5}
&&900&500&8\\\cline{3-5}
&&300&800&0.002\\
 \hline \hline
\end{tabular}
\caption{}
\label{table:LLP3bodydecay}
\end{subtable}
\caption{Decay length ($c\tau$) of $\chi^{+}$  in meters for typical masses of  $\psi$ and $S^{+}$ and at a fixed value of the coupling. All masses are in GeV.  Heavy neutrino mass is set to $m_N=30$  GeV.}
\label{table_chi2body}
\end{table}
Moving on to the case of $m_\chi<m_S+m_\psi$, the two-body decay is kinematically disallowed and the three body decay $\chi^{+}\rightarrow \psi N \ell^{+}$ is controlled by the product $y_1y_2$ and mediated by a virtual $S^*$, as shown in the Feynman diagram in Fig.~\ref{fd3}. In this case, $\chi^{+}$ can be lighter than 100 GeV as seen from Fig.~\ref{fig:case2_freezein1}, with the product of the couplings  mostly range between $10^{-10}$ and  $10^{-6}$. With conditions of mass splitting satisfied for enhanced co-annihilation, possibilities with larger couplings even up to $10^{-5}$ are not rare, and may even reach up  to $10^{-4}$ albeit with less probability, as presented in Table~\ref{table:finfout}.  In Fig.~\ref{fig_SS_ct} the decay length $c\tau$ is plotted against $m_\chi$ for exemplary values of the relevant coupling combination, keeping $m_\psi +m_N+m_\ell<m_\chi<m_\psi+m_S$ so as to focus on the 3-body decay. The dependence on the mass here is stronger than that in the 2-body decay case discussed above. For the entire mass range of $m_\chi=40$~GeV to large values of $m_\chi=1000$~GeV it is possible to  have $\chi^{+}$ decaying within the Inner Detector or the Calorimeters, or even decay outside the detectors, leaving a LLP signature. Table~\ref{table:LLP3bodydecay} shows the decay lengths for some specific benchmark points relevant to the LHC.
 
\subsection{Long-lived charged scalar, $S^{+}$}
\begin{figure}[H]
\begin{center}
\includegraphics[width=0.55 \textwidth]{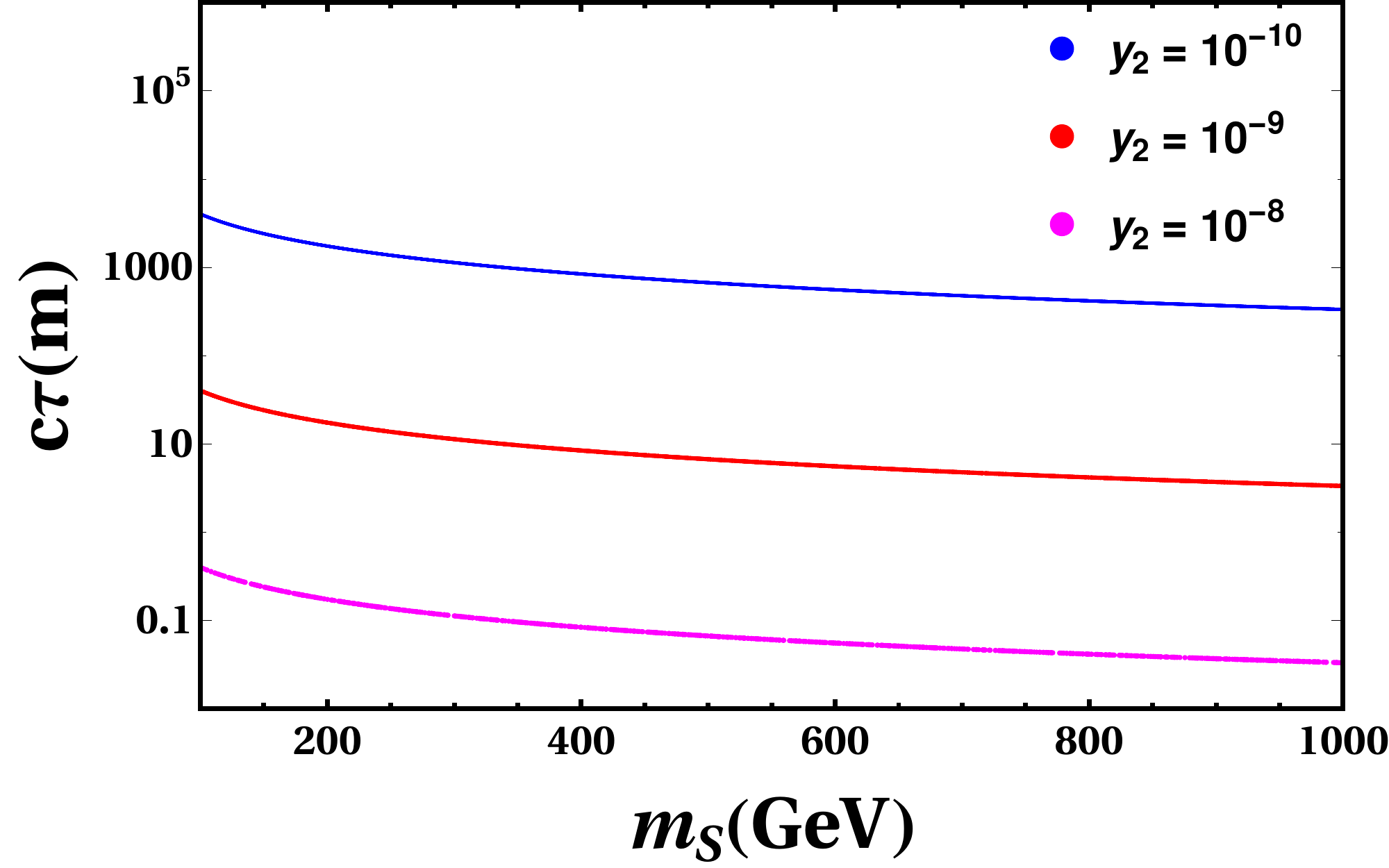}
\caption{Decay length vs.~mass of long lived particle ($S^{+}$) for different couplings. $m_{N}$ is fixed at 30 GeV.}
\label{S_decay_length} 
\end{center}
\end{figure}

Turning to the case of $S^{+}$, its decay length is decided purely by the strength of $y_2$. For the freeze-in solution of dark matter, we needed $y_1\lesssim 10^{-8}$ so as to be compatible with the observed relic density. The charged scalar, $S^{+}$ is mostly produced through gauge mediation, with a small contribution from the Higgs mediated process enabled by the quartic coupling $\lambda_1$.   When the coupling $y_1$ is of the order of unity, on the other hand, as required in the case of freeze-out mechanism to generate the dark matter relic density, $\chi^{+}$ would decay instantly as it is produced. It will decay to $S^{+}\psi$, with $S^{+}$ further decaying and $\psi$ missing. The cross section for this is given in Fig.~\ref{fig_SS_cs} against $m_S$ at 14 TeV LHC. The decay of $S^{+}$ is decided by the Yukawa coupling $y_2$. The decay width is given by
\begin{equation}
\Gamma_{S^{+} \rightarrow N l}=\frac{y_2^2}{16 \pi m_S^3}(m_S^2-m_N^2-m_l^2)\left[\left\{m_S^2-(m_N+m_l)^2\right\}\left\{m_S^2-(m_N-m_l)^2\right\}\right]^{\frac{1}{2}}\, .
\end{equation}
In Fig.~\ref{S_decay_length} we plot the decay length of $S^{+}$ against $m_S$ for specific choices of the relevant Yukawa coupling, $y_2$. Considering this to be of the order of $10^{-8}-10^{-10}$ makes it decay within the LHC detector, but with a measurable delay from its production so as to distinguish it from other particles which decay instantly. The signatures depend on the actual decay width. For $y_2\sim 10^{-8}-10^{-9}$ it leave a displaced vertex in the Inner Detector, whereas weaker couplings would make it decay within the Calorimeters or even in the muon detectors. Beyond $y_1\sim 10^{-10}$ will see $S^{+}$ escaping the LHC detectors without decaying, but leaving a clear charged particle track in all sectors. Similar to the case of $\chi^{+}$ discussed above, this can perhaps be distinguished from the muon tracks, owing to its heavy nature and slower speed.   It is also interesting that the mass splitting between $S^{+}$ and $N$ dictates the signatures of $S^{+}$ at the collider. If the splitting is small, a soft lepton will be produced in the final state, leaving disappearing tracks. On the other hand, if the splitting is large, then very likely a kink will be seen in the typical charged track signal.  The cross-sections and decay widths of $S^{+}$ for selected benchmark points are given in Table~\ref{table:collider_S} for some typical benchmark points which satisfy the relic density bound.

\begin{table}[h]
\begin{center}
 \resizebox{\linewidth}{!}{
\begin{tabular}{c |c |c |c |c |c |c |c |c }
\hline\hline
& ${\bm{m_{\psi}}}$ & ${\bm{m_S}}$ & ${\bm{m_{\chi}}}$ & ${\bm{y_1}}$ & ${\bm{\Omega h^2}}$ & ${\bm{\sigma_{p p\rightarrow S^+S^-}}}$ {\bf (pb)} & ${\bm{\Gamma_{S^{+} \rightarrow  N\ell}}}$ {\bf (GeV)} & ${\bm{c\,\tau}}$ {\bf (m)} \\ \hline\hline 
    
                                BP 1 & 56 & 165 & 473 & 7.53$\times 10^{-13}$  & 0.12& 0.0144 & 9.21$\times 10^{-18}$ & 21.5 \\
\hline
                                BP 2 & 101 & 77 & 613 & 1.18$\times 10^{-13}$  & 0.118&0.2441 & 3.3$\times 10^{-18}$ &  60.0 \\
\hline
                                BP 3 & 225 & 188 & 228 &  0.01 &  0.12& 0.0088 & 1.07$\times 10^{-17}$ & 18.5  \\
\hline
                                BP 4 & 150 & 127 & 154 &  0.05 &  0.119& 0.0376 & 6.76$\times 10^{-18}$ & 29.5  \\
\hline\hline
\end{tabular}}
\caption{Values of benchmark points considered in this paper for the production cross section (at 14 TeV LHC) and decay of the charged scalar $S^{+}$. DM relic density is achieved through either freeze-in (2 body) or freeze-out. All masses are in GeV. For all points, $m_{N}$ = 30 GeV and $y_{2}=10^{-9}$.}
\label{table:collider_S} 
\end{center}
\end{table}

\subsection{Long-lived neutral fermion, $N$}
The neutral heavy fermion, { $N$, decays through $N \nu h$ interaction, where $\nu$ is the SM neutrino and $h$ is the SM Higgs boson}. The strength of this interaction is decided by the Yukawa coupling $y_N$, which also decides the neutrino mass through the seesaw mechanism. As discussed in the introduction, this coupling is required to be $\sim 10^{-8}$ for $m_N$ considered in our discussion. In addition, for $m_N < 125$~GeV it cannot decay to an on-shell Higgs boson, further slowing down the decay.  Therefore, $N$ (of mass in the 100 GeV range or lighter) decays much beyond the LHC detectors, consequently leaving missing energy signature, like the SM neutrinos. Dedicated detectors like the MATHUSLA~\cite{Curtin:2018mvb} placed beyond the LHC detector complex could capture the signatures of such long lived particles. This surface detector is typically prepared to identify decay of LLP into a pair of charged particles, which are detected by its multi-layer trackers. However, notice that $N$ decays to neutrino and Higgs boson. With the missing energy, it will be difficult to correlate the event as originating from the interaction point. Different techniques like time of flight information may need to be made use of, requiring close talks between the CMS/ATLAS and MATHUSLA detectors. On the other hand, for $m_N \gtrsim 130$ GeV, $N$ can have decay length within the CMS/ATLAS detector, and can in principle be probed through the displaced vertex study. Again, the typical search strategies adopted in the case of displaced vertex studies of ATLAS and CMS will not work in  this case. It may be required to look for reconstructed Higgs (produced almost at rest, especially when $N$ is nearly degenerate with Higgs) originating from a displaced vertex along with associated missing transverse momentum with respect to the direction of the interaction point could be a possible signature.
\begin{table}[ht!]
\begin{center}
\begin{tabular}{c |c |c  }
\hline\hline
{\bm{$m_N$}} {\bf (GeV)}&{\bm{$\Gamma_N$}} {\bf (GeV)}&{\bm{$c\tau$}}{\bf  (m)}  \\ \hline
100& 1.53$\times 10^{-22}$&1.3$\times 10^{6}$\\ \hline
120& 7.53$\times 10^{-22}$& 2.8$\times 10^{5}$\\ \hline
128& 4.09$\times 10^{-19}$& 480.5\\ \hline
130& 1.1$\times 10^{-18}$& 179 \\ \hline
135& 4.1$\times 10^{-18}$& 48.05 \\ \hline
150& 2.09$\times 10^{-17}$& 9.42 \\ \hline
200& 1.1$\times 10^{-16}$& 1.79\\
\hline
\hline
\end{tabular}
\caption{{Decay width and decay length of $N$. $y_N=10^{-8}$ in all cases. It is to be noted that the order of $\Gamma_N$ changes substantially for $m_N \ge$ 125 GeV, which implies the onshell production of SM Higgs.}}
\label{Ndecay_table}
\end{center}
\end{table}
Although we have not considered heavier $N$ for our analysis, in order to get the same relic density, the increase in $m_N$ in the decay width of $\chi$ can be compensated by larger $y_1y_2$.

Finally, we may consider these search options in the proposed $e^+e^-$ colliders like CLIC with sufficient centre of mass energy to produce the charged particle, $\chi^{+}$ and $S^{+}$. The search strategies in this case would be very similar to that mentioned above in the case of the LHC. Notice that the production being electroweak process, cross section in this case will be comparable with the LHC. On the other hand, the clean environment and fixed centre of mass energy can be of great advantage in the leptonic colliders.  We shall defer a detailed study to a future publication.

%
\section{Conclusion}
\label{Conclusion}
Providing a viable dark matter candidate, yet compatible with all the experimental observations is one of the major challenges of particle physics today. In the popular WIMP scenario, weakly interacting dark matter particles produced thermally in abundance is depleted with large enough annihilations to achieve the observed relic density as they decouple and freeze out. In most models of WIMP the basic interaction that leads to such annihilations is the same as that is relevant in the direct detection experiments based on their elastic scattering from nuclei. Presently, the direct detection experiments have constrained these couplings to the extent that it becomes quite difficult to accommodate the required annihilation cross section. This has led many researchers to favour an alternate mechanism, whereby very feebly interacting massive particles are slowly produced to build up the relic density of the dark matter. In this FIMP scenario, usually, a thermally produced partner particle decays very slowly to the dark matter particle. 

In this work we consider a gauge singlet fermionic FIMP dark matter, $\psi$  and an isospin singlet charged fermion, $\chi^{+}$ as its partner particle. The stability of the dark matter particle is ensured by imposing a $Z_2$ symmetry under which both $\chi$ and $\psi$ are odd, while all other particles are even.  Gauge produced  $\chi^{+}$ decays through its Yukawa coupling ($y_1$) to $\psi$ and another newly introduced isospin singlet charged scalar $S^{+}$.  For very small values of this Yukawa coupling, the dark matter density will be slowly built up. Our analysis shows that for $10^{-13}\le y_1\le 10^{-8}$ a wide range of DM mass 1 GeV $\le m_\psi\le 600$ GeV is viable for partner mass of 100~- 1000 GeV.  The presence of $S^{+}$ and the Yukawa coupling ($y_2$) enabling its decay to a heavy neutral fermion and the charged SM leptons, add an entirely new dimension to the dynamics compared to what is usually discussed in the literature. The signatures here include a slowly decaying long-lived particle, as well as the dark matter particle itself. Under favourable conditions, the partner particles can be produced in sufficient amounts at colliders like the LHC, which then will decay away from the interaction point, but within the detector, leaving distinct signatures compared to many exotic particle searches. The typical coupling (relevant to the decay) required for this is $\sim 10^{-9}$, which in the usual FIMP scenarios discussed in the literature require very light dark matter candidates (keV - MeV). In the present scenario we discussed, the presence of $y_2$, which can be  independent of the dark matter considerations makes it possible to have both $\chi^{+}$ and $S^{+}$ to be a long lived particles  that could be searched at the LHC, for a large range of its mass. When the kinematic condition  $m_\chi>m_\psi+m_S$ is not met, $\chi^{+}$ decays through off-shell $S^{+}$ to $N\ell^{+}\psi$. This decay is controlled by a combination of $y_1$ and $y_2$. With $y_1$ within a restricted range, $y_2$ can be arranged so that the delayed decay of $\chi^+$ to the DM candidate $\psi$ adds significantly to is number density. This is a new feature of the dark matter dynamics that is comparatively less explored  in the literature. This intermingling of slow production leading to freeze-in and annihilation followed by freeze-out opens up a large region of the parameter space, otherwise unavailable when these mechanisms are considered separately. We have discussed the details of this in Section~\ref{freezeincase2}, where the importance of annihilation process is clearly established. In sharp contrast to the usual freeze-in plus LLP scenario, the DM mass range that is compatible in the present case is from a few GeV to TeV. While the usual HSCP scenario arises in certain parameter regions, there are other distinct features requiring modified search strategy at LHC/MATHUSLA detectors as summarised below.
\begin{enumerate}
\item Having two charged $LLP$'s, it is possible to have $LLP1\to LLP2 + \slash \!\!\!\! E $, with $LLP2$ decaying outside the detector. With large enough missing energy, this can lead to a HSCP track with a kink at the first decay vertex.

\item  $LLP1\to LLP2 + MET  \to SM + MET$. Within favourable parameter regions, this can lead to sufficiently large disappearing track, with a kink at the first decay vertex.

\item The neutral $LLP$ decay to Higgs boson and neutrino ($N\to h\nu$). With $h$ further decaying to $b\bar b$, this could lead to two jets with displaced vertex, and having invariant mass around Higgs mass, within ATLAS/CMS or MATHUSLA.  The reconstructed Higgs momentum will not point to the interaction point owing to large associated missing energy of the undetected neutrino.
\end{enumerate}
We have explored various possibilities, and show that for $m_\psi$ starting from 1~GeV all the way up to 1000~GeV and beyond, almost the entire range of theoretically allowed coupling values and a large range of masses of other particles are compatible with the observations.  The collider phenomenology in this region of parameter space may not feature any long-lived particle. On the other hand, presence of two exotic charged particles, one fermion and one scalar will add rich phenomenology, which will be explored in detail in a different work. For easy reference, we have summarised the conclusions in a qualitative way for different ranges of the relevant couplings in Table~\ref{table:summary}.

\begin{table}[H]
   \centering
  \resizebox{\linewidth}{!}{
   \begin{tabular}{c |c |c |c |c |c }
  \hline\hline
 \multirow{2}*{\bf DM scenario} & \multicolumn{3}{c|}{{\bf couplings}} & \multirow{2}*{\bf LLP} & \multirow{2}*{\bf collider signature}\\ \cline{2-4} 
                                & {\bf $y_1$} & {\bf $y_2$} & {\bf $y_1.y_2$} &  & \\ \hline \hline
  {\it Freeze-in via} & \multirow{2}*{$10^{-13} \lesssim y_1 \lesssim 10^{-9}$} & \multirow{2}*{$10^{-10} \lesssim y_2 \lesssim 10^{-7}$} & \multirow{2}*{small} & $\chi^{+}$ & charged track \\ \cline{5-6}
  {\it 2 body decay} & & & &  $S^{+}$ & displaced vertex, charged track \\ \hline
  {\it Freeze-in via} & \multirow{2}*{$\lesssim 10^{-8}$} & \multirow{2}*{$10^{-3} \lesssim y_2 \lesssim 1$} & \multirow{2}*{$10^{-11} \lesssim y_1y_2 \lesssim 10^{-8}$} & $\chi^{+}$ & charged track \\
  {\it 3 body decay} & & & & & \\ \hline
  {\it thermal production} & \multirow{2}*{$10^{-8} \lesssim y_1 \lesssim 10^{-7}$}  & \multirow{2}*{$10^{-3} \lesssim y_2 \lesssim 1$ } & \multirow{2}*{$10^{-11} \lesssim y_1y_2 \lesssim 10^{-7}$} & $\chi^{+}$ & charged track \\
  {\it + non-thermal production} & & & &  &  \\
  {\it through $\chi^+$ decay (dominant)}& & & &  &  \\ \hline
  \multirow{2}*{\it Freeze-out} & \multirow{2}*{$\sim 1$} & \multirow{2}*{$ \lesssim 10^{-7}$} & \multirow{2}*{ $\lesssim 10^{-4}$} & $\chi^{+}$ & charged track, displaced vertex \\ \cline{5-6}
  & & & & $S^{+}$ & displaced vertex,charged track \\ \hline
  {\it Freeze-out} & $\sim 1$ & $\sim 1$ & $\sim 1$ & - & - \\ \hline
 \hline
  \end{tabular}}
  \label{summary_table}   
  \caption{Table showing the possible DM scenarios and the limits on the coupling from relic density and collider requirements. LLP possibilities and signatures are also mentioned.}
  \label{table:summary}
\end{table}

Finally, we point the attention of the reader to the presence of heavy neutral fermion inducing type-I seesaw mechanism providing a way to generate light neutrino mass within the same framework, as an added feature of the model.

\appendix
\section{Boltzmann Equations}
\label{appendix:BE}
The dark matter particle $\psi$ when produced with slow decay of $\chi^+$, but with a rate that results in sufficiently large number density of $\psi$ before freezing-in presents a novel feature as discussed towards the end of Section~\ref{freezeincase2}. With sufficiently large value of the relevant coupling, $y_1$, the thermal production and pair annihilation become significant, leading to freeze-out. The two limiting cases are {\em (i)} with $y_1$ is small enough so that only the production through $\chi^+$ decay is relevant, and {\em (ii)} with $y_1$ large enough so that the thermal production and subsequent freeze-out scenario are relevant. The Boltzmann Equation controlling the yield in this case depends on the annihilation cross section of $\psi$ along with the decay rate of $\chi^+$ as given by
\begin{gather}
\frac{dY_{\psi}}{dx} = \sqrt{\frac{45}{4\pi^3G}}\frac{x}{m^2_{\psi}}\frac{\sqrt{g_*(T)}}{g_s(T)}(\langle\Gamma_{\chi \rightarrow \psi X}\rangle Y_{\chi_{eq}}) \nonumber -\sqrt{\frac{\pi}{4 \pi G}}\frac{m_{\psi}}{x^2}\sqrt{g_*(T)}\langle \sigma v \rangle_{\chi}(Y_{\psi}^2-Y_{\psi_{eq}}^2),
\end{gather}
with $Y_{\chi_{eq}}$ and $Y_{\psi_{eq}}$ are the equilibrium number densities for $\chi^{+}$ and $\psi$ respectively, $x=m_{\psi}/T$, where $T$ is the photon temperature.  Notice that in the standard freeze-in case $x$ is defined in terms of the mass of the decaying particle ($\chi^+$ in our case). However, what is relevant in the present case is the final freeze-out, and therefore we have considered $x=\frac{m_\psi}{T}$, appropriately redefining other factors.
\begin{equation}
\sqrt{g_*(T)}\simeq\frac{g_s(T)}{\sqrt{g_{\rho}(T)}} \nonumber
\end{equation}
where $g_s$ and $g_{\rho}$ are the degrees of freedom corresponding to entropy and energy density of the Universe.
\begin{equation}
\langle \Gamma_{\chi^{+} \rightarrow \psi X} \rangle = \Gamma_{\chi^{+} \rightarrow \psi X} \frac{K_1(x^{'})}{K_2(x^{'})} \nonumber 
\end{equation}
$x^{'}=m_{\chi}/T$, $K_1$ and $K_2$ are modified Bessel functions and $\Gamma_{\chi^{+} \rightarrow \psi X}$ represents the decay width of $\chi^{+}$ to $\psi$ and X, where X denotes $S^{+}$ in case of 2 body and N and $l^{+}$ for 3 body decay of $\chi^{+}$.
$\langle \sigma v \rangle_{\chi^{+}}$ denotes the thermal average of all possible pair annihilation and coannihilation cross-sections of $\psi$.

The relic density is related to the yield through the standard relation,
\begin{gather}
\Omega_{\psi}h^2 = 2.75 \times 10^8 m_{\psi} Y_{\psi}(T_0),
\end{gather}
where $T_0$ is the present temperature of the universe.

\section{Constraint from $H\to\gamma\gamma$} \label{sec:hgg}
The Higgs boson has a direct quartic coupling with the new charged scalar, $S^+$, 
\begin{equation}
{\cal L} \ni \lambda_1 S^\dagger S H^\dagger H.
\end{equation}
The trilinear coupling arising from this after the electroweak symmetry breaking results in $S^+$ contribution to $H\to \gamma \gamma$ at the leading order through the triangle diagram.
Including this contribution from $S^+$, the diphoton decay width is given by \cite{Gunion:1989we, Djouadi:2005gi}
\begin{equation}
\Gamma (H\to \gamma\gamma)=\frac{\alpha^2 m_H^3}{256~\pi^3v^2}~\left| F_W+\sum_fN_{cf}Q_f^2~F_f+\lambda_1v~F_S\right|^2,
\end{equation}
where
\begin{eqnarray}
F_W&=&2+3\tau_W\left[1+\left(2-\tau_W\right)f(\tau_W)\right],\\
F_f&=&-2\tau_f\left[ 1+(1-\tau_f)f(\tau_f)\right],\\
F_S&=&\tau_S\left[1-\tau_Sf(\tau_S)\right],
\end{eqnarray}
with $\tau_i=\frac{4m_i^2}{m_h^2}$ and
\begin{eqnarray}
f(\tau)=\left\{\begin{matrix}{\rm ArcSin}^2(1/\sqrt{\tau}),~~~~~~{\rm for}~~\tau \ge 1\\
-\frac{1}{4}\ln\left(\frac{1+\sqrt{1-\tau}}{1-\sqrt{1-\tau}}\right)-i\pi,~~~~~~{\rm for}~~\tau <1.
\end{matrix} \right.
\end{eqnarray}
With $m_S = 150$ GeV, we have 
\begin{equation}
\Gamma(H\to \gamma\gamma)\sim 0.25\ ({\rm keV})~\left( 6.54-90.5~\lambda_1 \right)^2.
\label{eq:hgg}
\end{equation}

\section{Constraint from Invisible Higgs decay}\label{sec:hpsipsi}
Higgs boson does not have direct coupling to the dark matter, $\psi$. 
\begin{figure}[ht!]
\begin{center}
\includegraphics[width=0.2\linewidth]{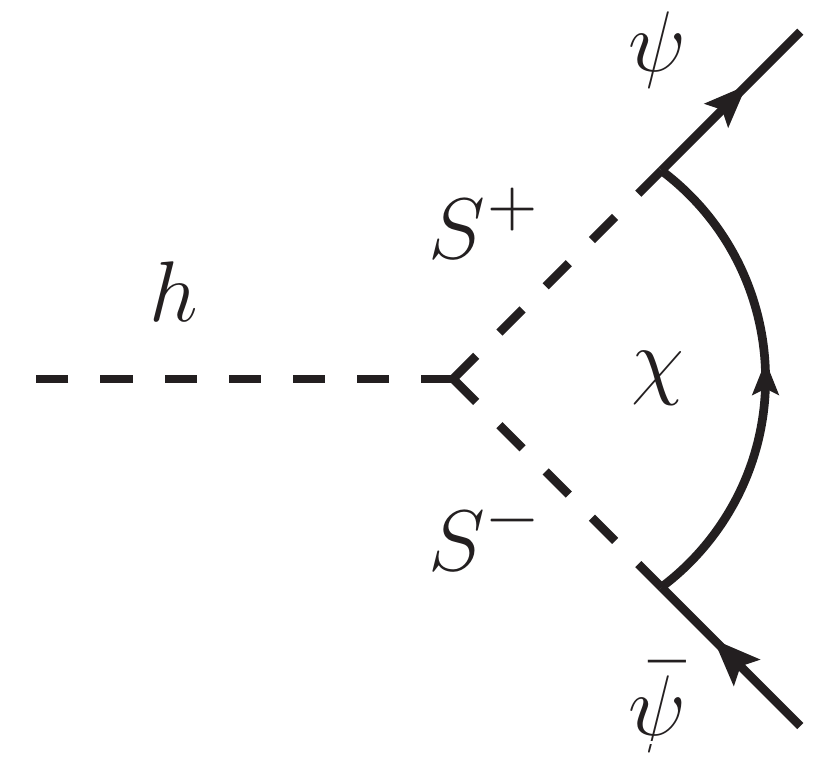}
\caption{Feynman diagram contributing to the invisible Higgs decay through $H\to \psi\psi$.}
\label{fig:hpsipsi}
\end{center}
\end{figure}
However, $H\to \psi\psi$ is possible through the one loop diagram as in Fig.~\ref{fig:hpsipsi}. This  contribution to the  invisible decay of the Higgs is given by \cite{Herrero-Garcia:2018koq}

\begin{eqnarray}
\small
  \Gamma (H \to \psi \psi)
&  =&
  \frac{m_H}{2\,\pi}\,\left[1 - \frac{4m^2_\psi}{m^2_H}\right]^{3/2}\,|g_V|^2,
\end{eqnarray}
with
\begin{eqnarray}
\small
  g_V&
  =&
  \frac{\lambda_1\,y^2_1\,\,v}{32\,\pi^2}\,\left[2\,m_\psi\,C_1(m^2_\psi, m^2_H, m^2_\psi, m_\chi, m_S, m_S) - m_\chi\,C_0(m^2_\psi, m^2_H, m^2_\psi, m_\chi, m_S, m_S)\right], \nonumber
\end{eqnarray}
where $C_i$ are the relevant Passarino-Veltman loop functions. With all new particle masses around GeV-TeV scale, $\Gamma(H \to \psi \bar\psi) \sim {\cal O}(10^{-6})$ for  the couplings $y_1 \sim 1$ and $\lambda_1 \sim 0.01$  which is negligible compared to the standard contributions. 

\section{Direct Detection Constraints}\label{sec:dd}
\begin{figure}[h]
\vskip 4mm
\begin{center}
\includegraphics[width=0.5\linewidth]{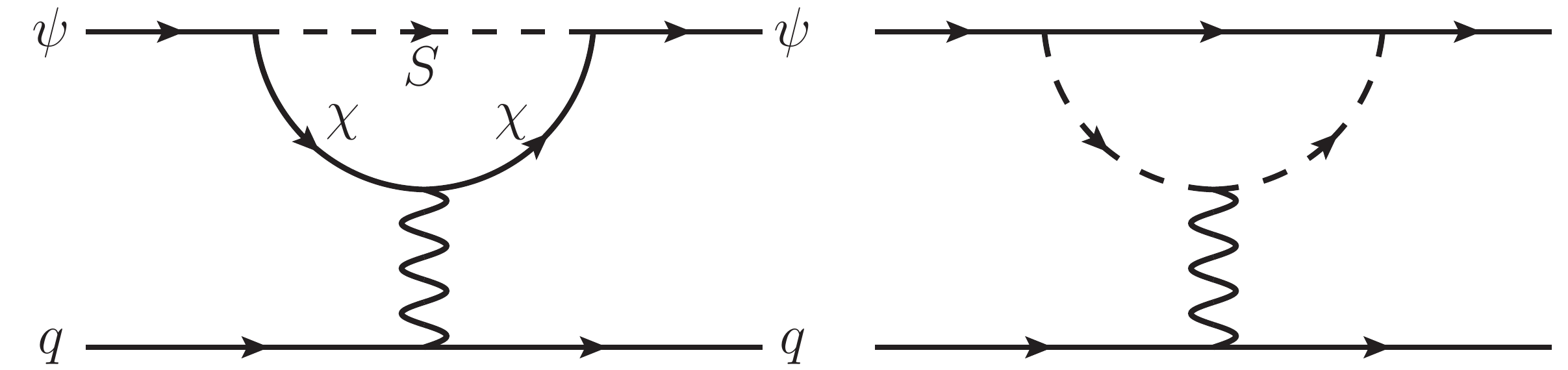}
\vskip 2mm
\caption{One-loop diagrams that contribute to dark matter - nucleon scattering. Notice that there are no tree-level contributions.}
\label{fig:dd}
\end{center}
\end{figure}

There are two sets of Feynman diagrams with photon and $Z$ mediation, as in Fig.~\ref{fig:dd}. 
The spin-independent direct detection cross section is given by~\cite{Chao:2019lhb}
\begin{align}
  \sigma^{\psi\chi}_{\rm SI}
  =
  \frac{16}{\pi}\,\frac{m^3_\psi m^3_p}{m_\psi + m_p}\,\,|c^q_{VV}|^2 \,,
\end{align}
where $m_p$ is the proton mass. The relevant Wilson coefficients are given by
\begin{align}
  c^q_{VV}
  =&
  - \frac{\alpha_{\rm em}}{24\,\pi\,x^4_\psi}\,\frac{1}{m^2_S}\,Q\,Q_q\,y^2_1
  \bigg[\Big( - 3\,x^6_\psi + 6\,x^5_\psi\,x_\chi + 12\,x_\psi\,x_\chi\,(1 - x^2_\chi)^2 + 8\,(1 - x^2_\chi)^3 + 2\,x^4_\psi\,(5 + x^2_\chi)
  \nl
  &
  - 6\,x^3_\psi\,x_\chi\,(1 + 3\,x^2_\chi) - 3\,x^2_\psi\,(5 - 2\,x^2_\chi - 3\,x^4_\chi)\Big)\,\frac{g (x_\psi, x_\chi)}{1 - (x_\psi - x_\chi)^2}
  + \frac{2\,x^2_\psi\,(4 - 3\,x^2_\psi + 6\,x_\psi\,x_\chi - 4\,x^2_\chi)}{1 - (x_\psi - x_\chi)^2}
  \nl
  &
  + (8 + x^2_\psi - 4\,x_\psi\,x_\chi - 8\,x^2_\chi)\,\ln x_\chi \bigg]\,.
\end{align}
The loop function in terms of the mass ratios $ x_\psi \equiv \frac{m_\psi}{m_S}$ and 
$  x_\chi \equiv \frac{m_\chi}{m_S}$ is given by
\begin{align}
  g (x_\psi, x_\chi)
  =
  \frac{\ln\left(1 - x^2_\psi + x^2_\chi + \sqrt{x^4_\psi + (1 - x^2_\chi)^2 - 2 x^2_\psi (1 + x^2_\chi)}\right) - \ln(2 x_\chi)}{\sqrt{x^4_\psi + (1 - x^2_\chi)^2 - 2 x^2_\psi (1 + x^2_\chi)}}\,.
\end{align}

\acknowledgments

PP and SC thank DST-SERB, India for the research project grant EMR/2015/000333, and the DSTFIST grant SR/FST/PSIl-020/2009 for offering the computing resources needed by this work. SC would like to acknowledge fruitful discussions with Amit Dutta Banik and Rashidul Islam and thank MHRD, Govt. of India for research fellowship. We also thank the British Council, India for sponsoring the visit of VM to IIT Guwahati during which the collaboration was initiated.



\bibliography{frlng}
\bibliographystyle{jhep}
\end{document}